\documentclass[useAMS,usenatbib]{mn2e}

\usepackage{graphicx}
\usepackage{amsmath}
\usepackage{amssymb}
\usepackage{bm}
\usepackage{ulem}

\defcitealias{css12}{Paper~I}
\defcitealias{ms91}{MS91}
\defcitealias{sm93}{SM93}
\defcitealias{rss88}{RSS88}
\shortcites{Ketal03}

\newcommand{\be}{\begin{equation}}
\newcommand{\ee}{\end{equation}}

\newcommand{\bsplit}{\begin{split}}
\newcommand{\Exp}[1]{{\rm e}^{#1}}
\newcommand{\del}{\partial}
\newcommand{\Del}{{\nabla}}
\newcommand{\alp}{\alpha}
\newcommand{\Gam}{\Gamma}
\newcommand{\eps}{\epsilon}

\newcommand{\Emf}{\bm{\mathcal{E}}}

\newcommand{\mean}[1]{\overline{#1}}
\newcommand{\meanv}[1]{\overline{\bm #1}}

\newcommand{\corr}{_\mathrm{c}}						
\newcommand{\corot}{_\mathrm{cor}}						
\newcommand{\Dyn}{_{D}}									
\newcommand{\f}{_\mathrm{0}}					   	
\newcommand{\kin}{_\mathrm{k}}			   		
\newcommand{\magn}{_\mathrm{m}}			   		
\newcommand{\turb}{_\mathrm{t}}			   		
\newcommand{\crit}{\mathrm{cr}}			   		
\newcommand{\const}{\mathrm{const}}			   		
\newcommand{\ma}{_\mathrm{max}}			   		
\newcommand{\dd}{\mathrm{d}}			   		
\newcommand{\diff}{_\mathrm{d}}			   		
\newcommand{\real}{_\mathrm{R}}			   		
\newcommand{\imag}{_\mathrm{I}}			   		

\newcommand{\cro}{\times}

\newcommand{\mbr}{\mean{B}_r}
\newcommand{\mbp}{\mean{B}_\phi}
\newcommand{\mbz}{\mean{B}_z}

\newcommand{\alphatilde}{\widetilde{\alpha}}
\newcommand{\etatilde}{\widetilde{\eta}}

\newcommand{\omegatilde}{\widetilde{\omega}}
\newcommand{\omtilde}{\omegatilde}

\newcommand{\atilde}{\widetilde{a}}
\newcommand{\btilde}{\widetilde{b}}
\newcommand{\Atilde}{\widetilde{A}}
\newcommand{\Btilde}{\widetilde{B}}
\newcommand{\betatilde}{\widetilde{\beta}}

\newcommand\bgreek[1]{ \mathchoice
    {\hbox{\boldmath$\displaystyle{#1}$\unboldmath}}%
    {\hbox{\boldmath$\textstyle{#1}$\unboldmath}}%
    {\hbox{\boldmath$\scriptstyle{#1}$\unboldmath}}%
    {\hbox{\boldmath$\scriptscriptstyle{#1}$\unboldmath}}}

  \newcommand{\cm}{\,{\rm cm}}

  \newcommand{\kms}{\,{\rm km\,s^{-1}}}

  \newcommand{\kpc}{\,{\rm kpc}}
  
  \newcommand{\Mpc}{\,{\rm Mpc}}
  \newcommand{\Myr}{\,{\rm Myr}}
  \newcommand{\Gyr}{\,{\rm Gyr}}
  
  \newcommand{\mkG}{\,\mu{\rm G}}
  \newcommand{\nG}{\,{\rm nG}}

\title[Galactic spiral patterns and dynamo action II]{Galactic spiral patterns and dynamo action II:\\ Asymptotic solutions}
\author[L. Chamandy, K. Subramanian and A. Shukurov]{Luke Chamandy$^{1}$, Kandaswamy Subramanian$^{1}$ \& Anvar Shukurov$^{2}$
\thanks{E-mail: luke@iucaa.ernet.in (LC); kandu@iucaa.ernet.in (KS); anvar.shukurov@newcastle.ac.uk (AS)}\\
$^{1}$Inter-University Centre for Astronomy and Astrophysics, Post Bag 4, Ganeshkhind, Pune 411007, India\\
$^{2}$School of Mathematics \& Statistics, Newcastle University, Newcastle upon Tyne NE1 7RU}

\begin{document}

\date{Submitted to MNRAS}

\pagerange{\pageref{firstpage}--\pageref{lastpage}} \pubyear{2013}

\maketitle

\label{firstpage}

\begin{abstract}
The exploration of mean-field galactic dynamos affected by a galactic spiral pattern, 
begun in \citet[][hereafter \citetalias{css12}]{css12} with numerical simulations, 
is continued here with an asymptotic solution.
The mean-field dynamo model used generalizes the standard theory 
to include the delayed response of the mean electromotive force to variations of the mean magnetic 
field and turbulence parameters (the temporal non-locality, or $\tau$ effect).
The effect of the spiral pattern on the dynamo considered is the enhancement of the 
$\alpha$-effect in spiral-shaped regions  
(which may overlap the gaseous spiral arms or be located in the interarm regions). 
The axisymmetric and enslaved non-axisymmetric modes of the mean magnetic field 
are studied semi-analytically to clarify and strengthen the numerical results. 
Good qualitative agreement is obtained between the asymptotic solution 
and numerical solutions of \citetalias{css12} for a 
global, rigidly rotating material spiral (density wave).
At all galactocentric distances except for the co-rotation radius, we
find magnetic arms displaced
in azimuth
from the $\alpha$-arms,
so that the ridges of magnetic field strength are more tightly wound than the $\alpha$-arms. 
Moreover, the effect of a finite dynamo relaxation time $\tau$
(related to the turbulence correlation time)
is to phase-shift the magnetic arms in the direction opposite to the galactic rotation
even at the co-rotation radius.
This mechanism can be used to explain the 
phase shifts between magnetic and material arms observed in some spiral galaxies.
\end{abstract}


\section{Introduction}
\label{sec:intro}
Disc galaxies typically contain regular (or large-scale or mean) magnetic fields of 
1--$10\mkG$ in strength.
In many cases, their non-axisymmetric components can be 
as strong as the axisymmetric part \citep{f10}.
The non-axisymmetric part of the field takes the form of `magnetic spiral arms',
wherein the regular field is enhanced \citep{beetal96, sh05, be12}.
Magnetic arms cannot be 
accounted for 
if the underlying disc is assumed 
to be axisymmetric \citep[][hereafter \citetalias{css12}]{css12} and they appear to be 
related (though in a non-trivial way) to the material (gaseous) spiral arms
\citep{fricketal00}.
Hence the need to develop a theory which relates the non-axisymmetry of the regular magnetic field 
to that of the underlying disc.
This was the general aim of \citetalias{css12}, where we approached the problem from a numerical 
standpoint 
and considered not only the linear growth of the field (kinematic regime) but also the non-linear 
saturation phase.
Here we take an analytical approach and therefore restrict the investigation 
mostly to the kinematic regime,
although a nonlinear extension is also suggested.
We are motivated, in part, by insights that were provided by analytical treatments in the past 
(\citealt[][hereafter \citetalias{rss88}]{rss88}, for the axisymmetric case and \citealt[][hereafter \citetalias{ms91}]{ms91}, for the non-axisymmetric case). 
The latter work was focused on how the $|m|=1$ (bisymmetric) azimuthal modes can be generated in a galactic disc with a two-arm spiral pattern.
Numerical models (e.g. \citealt{mo98,rohdeetal99}; \citetalias{css12}) show, however, 
that $m=0$ (axisymmetric) and $|m|=2$ (quadrisymmetric) modes tend to be more prevalent in such discs (or $m=0$ and $|m|=n$ for discs with $n$ spiral arms).
An analytical model that can explain such modes is therefore needed.

Our aim in this paper is 
 a systematic interpretation of the key numerical results of 
\citetalias{css12} 
by solving a suitably simplified but essentially the same
mean-field dynamo equations with an approximate semi-analytical method. 
Here we consider both axially symmetric and enslaved non-axisymmetric modes of a kinematic dynamo. 
The enslaved modes are those that have the same growth rate as the leading axisymmetric mode; 
they occur because of deviations of the galactic disc from axial symmetry, 
e.g., due to a spiral pattern in the interstellar gas. 
For a two-armed spiral, the enslaved modes include the $|m|=2$ components
that co-rotate with the spiral pattern, 
as well as other, weaker, even-$m$ corotating modes. 
We leave the $|m|=1$ modes and other odd-$m$ corotating modes for a future work.
Our model differs from earlier analytical study of \citetalias{ms91} in that here we:
\begin{enumerate}
\item[(i)] provide an asymptotic treatment of the $m=0$ and 
even-$m$ modes forced by a rigidly rotating two-arm spiral; \\
\item[(ii)] incorporate the minimal-$\tau$ approximation (MTA) and explore the effects of a finite dynamo relaxation time.
\end{enumerate}

The plan of the paper is as follows.  The basic equations and mathematical approach
are summarized in Section~\ref{sec:equations}.
In Section~\ref{sec:semi-analytic}, we present an asymptotic solution for the axisymmetric 
and enslaved non-axisymmetric magnetic modes which inhabit a disc containing a global, steady, 
rigidly rotating spiral density wave. We then compare our results with numerical solutions 
in Section~\ref{sec:comparison},
including the saturated states.
Conclusions and discussion for both \citetalias{css12} and this paper can be found in 
Sections~\ref{sec:conclusion} and \ref{disc}.

\section{The generalized mean-field dynamo equation}
\label{sec:equations}
The basic equation solved here is a generalization of the standard mean-field dynamo equation,
and is derived in \citetalias{css12},
\begin{equation}
\label{telegraph}
\begin{split}
\tau \frac{\partial^2 \meanv{B}}{\partial t^2}+\frac{\partial \meanv{B}}{\partial t}&=
\tau \nabla \times\left( \meanv{U} \times \frac{\partial \meanv{B}}{\partial t} \right)\\
&\quad+{\bf \nabla} \times 
\left( \meanv{U} \times \meanv{B} + c_\tau\alpha \meanv{B} \right)
+ c_\tau\eta_t \nabla^2 \meanv{B}.
\end{split}
\end{equation}
Here we use the same (standard) notation as in \citetalias{css12}. We have neglected the terms 
$\tau\eta\Del^2({\del\meanv{B}}/{\del t})$ and $\eta\nabla^2 \meanv{B}$ 
on the right-hand side because of the high Ohmic conductivity of the interstellar gas, 
$\eta\ll\eta\turb$. Following the usual mean-field approach, the velocity and magnetic 
fields,  $\bm{U}$ and $\bm{B}$, have each been written as the sum of an average part and a random part,
\[
\bm{B}=\meanv{B}+\bm{b} \quad {\rm and} \quad \bm{U}=\meanv{U}+\bm{u},
\]
where overbar represents the ensemble average but for practical purposes spatial averaging 
over scales larger than the turbulent scale but smaller than the system size can be used
\citep{G92,E12,FSFSM12}.

Equation~\eqref{telegraph} arises when the response time $\tau$ of the mean electromotive force (emf) 
$\Emf=\overline{\bm{u}\times\bm{b}}$ to changes in the mean field or small-scale turbulence is not negligible \citep[e.g.][]{rheinhardtb12}.
The $\tau$ effect can have various interesting implications (see, e.g., \citet{BKM04} and \citealt{hubbardb09}, 
the latter of which also contains a brief review of applications from the literature).
A convenient approach to allow for this effect
is the minimal-$\tau$ approximation (MTA) \citep{rogachevskiikleeorin00, blackmanfield02, bs05a},
which leads to a dynamical equation for $\Emf$:
\begin{equation}
\label{emf}
\frac{\partial \Emf}{\partial t}=\frac{1}{\tau\corr}\left(\alpha\meanv{B}-\eta_t \bm{\nabla}\times\meanv{B}\right)-\frac{\Emf}{\tau}.
\end{equation}
In the kinematic limit, and for isotropic and homogeneous turbulence, the turbulent transport coefficients are given by
\begin{equation}
\alpha = -\tfrac{1}{3}\tau\corr\overline{\bm{u}\cdot{\bm{\Del\cro u}}}, 
\quad  
\eta\turb=\tfrac{1}{3}\tau\corr\overline{\bm{u}^2},
\end{equation}
with $\tau\corr$ the correlation time of the turbulence and $\tau$ the relaxation time,   
found in numerical simulations to be close to $\tau\corr$ \citep{bs05a,bs05b,bs07}. 
We retain $c_\tau=\tau/\tau\corr$ as a free parameter, but set it to unity in
illustrative examples. For the sake of simplicity, we take $\tau$ to be independent of scale and 
constant in space and time. 
Eq.~\eqref{emf}, when combined with the mean-field induction equation, leads to Eq.~\eqref{telegraph}.

The more standard mean-field dynamo equation
\be
\label{mean_induction}
\frac{\del\meanv{B}}{\del t} =\Del\cro(\meanv{U}\cro\meanv{B}+\alpha\meanv{B})+\eta\turb\Del^2\meanv{B},
\ee
is recovered from Eq.~\eqref{telegraph} in the limit $\tau\rightarrow0$ with $c_\tau=1$.
Below, we refer to the approach leading to Eq.~\eqref{mean_induction} as 
the `standard theory'.

\section{Asymptotic solutions for galactic dynamos}
\label{sec:semi-analytic}

Many disc galaxies have a spiral structure  
which causes deviations from axial symmetry in both turbulent and regular gas flows,
leading to non-axisymmetric magnetic fields. Here we follow \citetalias{ms91} by assuming 
that the $\alpha$ effect is modulated by the spiral pattern. 
The nature of such a modulation is largely unexplored \citep{sh98,SS98},
but it is reasonable to assume that $\alpha$ is enhanced along a spiral perhaps overlapping with the gas spiral.
The spiral patterns observed in the light of young stars are believed to represent a global
density wave rotating at a fixed angular frequency, or possibly a superposition of such waves, 
with differing azimuthal symmetries and angular frequencies \citep[e.g.][]{comparettaquillen12,roskaretal12}.
At least in some galaxies, they can be transient features \citep[e.g.][]{dobbsetal10, sellwood11, quillenetal11, wadaetal11, grandetal12} 
whose effect on the mean-field dynamo is discussed in \citetalias{css12}. 
In this paper, we consider an enduring spiral with $n$ arms and a constant pattern speed $\Omega$.

\subsection{Basic equations}
Since we are interested in non-axisymmetric magnetic fields that co-rotate 
with the spiral pattern, it is preferable to work in the corotating frame
where the dynamo forcing  is independent of time. Following \citetalias{ms91}, 
we carry out a coordinate transformation from the inertial cylindrical frame $\Sigma=(r,\phi,z,t)$ 
(with disc rotation axis as the $z$-axis)
to the frame $\widetilde{\Sigma}=(\widetilde{r},\widetilde{\phi},\widetilde{z},\widetilde{t})$ 
rotating with the pattern angular velocity $\Omega$ (assumed constant in both time and position):
\begin{equation}
\begin{split}
\label{coord_transf}
\widetilde{\phi}=\phi-\Omega t,\quad \widetilde{r}=r,\quad \widetilde{z}=z, \quad \widetilde{t}=t,\\
\widetilde{\mean{B}}_r=\mbr, \quad \widetilde{\mean{B}}_\phi=\mbp, \quad \widetilde{\mean{B}}_z=\mbz.
\end{split}
\end{equation}
so that
\begin{equation}
\label{deriv_transf}
\left(\frac{\partial}{\partial\phi}\right)_t=\left(\frac{\partial}{\partial\widetilde{\phi}}\right)_{\widetilde{t}}, \quad \left(\frac{\partial}{\partial t}\right)_\phi=\left(\frac{\partial}{\partial \widetilde{t}}\right)_{\widetilde{\phi}}-\Omega\left(\frac{\partial}{\partial\widetilde{\phi}}\right)_{\widetilde{t}}.
\end{equation}
The coordinate transformation shifts the mean velocity, 
$\widetilde{\overline{U}}_\phi=\overline{U}_\phi-r\Omega$,
but the random velocity is left unchanged,
$\widetilde{\bm{u}}={\bm{u}}$, as are both the mean and random magnetic fields.
Therefore, $\widetilde{\alpha}=\alpha$ and $\widetilde{\eta}\turb=\eta\turb$. 

We use the thin-disc approximation, $\partial/\partial z \gg \partial/\partial r$ and, 
eventually, consider tightly wound magnetic spirals, 
$\partial/\partial r \gg r^{-1}\partial/\partial \phi$. 
Thus, we consider magnetic fields whose radial scale is asymptotically intermediate between the 
 scale height of the galactic disc (of order 
$0.5\kpc$) and its 
 radial scale length (of order $10\kpc$). 
We also adopt the $\alpha\omega$-dynamo approximation where the rate
of production of the azimuthal magnetic field by the $\alp$ effect
is negligible in comparison with the effect of the differential rotation. 
The mean velocity field is taken to be axisymmetric, purely azimuthal, and constant in time,
\begin{equation}
\meanv{U}=r\omega(r)\widehat{\bgreek{\phi}},
\end{equation}
where $\widehat{\bgreek{\phi}}$ is the unit azimuthal vector. 
In the rotating frame, the gas angular velocity is $\widetilde{\omega}(r)=\omega(r)-\Omega$. 
The magnitude of the rotational velocity shear is quantified with 
$G(r)= r\dd\omega/\dd r =r\dd\widetilde{\omega}/\dd r$. 
Applying Eqs.~(\ref{coord_transf}) and (\ref{deriv_transf}) to the $r$- and $\phi$-components of 
Eq.~(\ref{telegraph}), and dropping tildes for ease of notation (except for on $\omegatilde$), 
leads to the following equations in the corotating frame:
\begin{align}
\label{telegraph_r_corot}
\widehat{\mathcal{L}}\,
\mbr
=&
\nonumber\mbox{}-c_\tau\frac{\partial}{\partial z}(\alpha \mbp)\\
&+c_\tau\eta\turb\left(\widetilde{\nabla}^2\mbr+\frac{1}{r^2}\frac{\del^2\mbr}{\del\phi^2}-\frac{2}{r^2}\frac{\del\mbp}{\del\phi}\right),\\ 
\label{telegraph_phi_corot}
\widehat{\mathcal{L}}\,
\mbp
=&
\nonumber\left(1+\tau\frac{\partial}{\partial t}-\Omega\tau\frac{\partial}{\partial\phi}\right)(G\mbr)\\
&+c_\tau\eta\turb\left(\widetilde{\nabla}^2\mbp+\frac{1}{r^2}\frac{\del^2\mbp}{\del\phi^2}+\frac{2}{r^2}\frac{\del\mbr}{\del\phi}\right),
\end{align}
where
\[
\widehat{\mathcal{L}}=
\left(1+\tau\frac{\partial}{\partial t}-\Omega\tau\frac{\partial}{\partial\phi}\right)\left(\frac{\partial}{\partial t}+\widetilde{\omega}\frac{\partial}{\partial\phi}\right)
\]
and
\[
\widetilde{\nabla}^2 X=
\frac{\partial^2 X}{\partial z^2}+\frac{\partial}{\partial r}\left[\frac{1}{r}\frac{\partial}{\partial r}(rX)\right].
\]
These equations agree with Eqs.~(2.2) and (2.3) of \citetalias{ms91} in the limit $\tau\rightarrow 0$ 
with $c_\tau=1$.
Since, in the thin-disc approximation, both equations do not include $\mbz$, the equation for $\mbz$ can be replaced by the solenoidality condition,
\begin{equation}
\frac{1}{r}\frac{\partial}{\partial r}(r\mbr)+\frac{1}{r}\frac{\partial \mbp}{\partial \phi}+\frac{\partial \mbz}{\partial z}=0.
\end{equation}

We emphasize that the governing equations acquire additional terms when 
transformed to the rotating frame whenever $\tau$ is finite. 
On the contrary, the standard mean-field dynamo equation \eqref{mean_induction} 
does not change form under the transformation to a rotating frame.
This can be understood physically in the following way.
In the standard theory, $\Emf$ at time $t_0$ and position $\bf{x}$ depends on $\meanv{B}$ and the turbulence parameters at the same time and position.
This is not true in the MTA, where $\Emf(t_0,\bf{x})$ depends on the history of 
$\meanv{B}(t,\bf{x})$, $\alphatilde$ and $\etatilde\turb$ over, roughly, $t_0-\tau<t\leq t_0$. 
[Here $\alphatilde$ and $\etatilde\turb$ refer to the more general version of Eq.~\eqref{emf}, 
wherein we have made the standard assumption $\alphatilde\approx\alpha/\tau\corr$, $\etatilde\turb\approx\eta\turb/\tau\corr$ \citep{bs05a}.]
Thus, $\Emf(t_0,\bf{x})$ is affected by how fast variations in the field and ($\alpha$, $\eta\turb$) are sweeping past the position $\bf{x}$. 

\subsection{Approximate solution}
When the coefficients of Eqs.~\eqref{telegraph_r_corot} and \eqref{telegraph_phi_corot} depend on 
the azimuthal angle $\phi$ as $\exp(i m\phi)$ with certain $m$, 
their solutions can be represented in the form
\be
\label{Bsum}
\left(\!\!
\begin{array}{c}
\mbr\\
\mbp
\end{array}
\!\!\right)
=
\sum_{m=-\infty}^\infty
\left(\!\!
\begin{array}{c}
a_m(r,z)\\
b_m(r,z)
\end{array}
\!\!\right)
\exp{(im\phi+\Gamma t)},
\ee
so that Eqs.~\eqref{telegraph_r_corot} and \eqref{telegraph_phi_corot} reduce to
\begin{align}
\nonumber\displaystyle\sum_{m=-\infty}^\infty &\Exp{im\phi}\Biggl[\mathcal{L}a_m\Biggr.
\label{a_m}
+c_\tau\frac{\del}{\del z}(\alpha b_m)\\
&\Biggl.-c_\tau\eta\turb\left( \widetilde{\nabla}^2
a_m-m^2\frac{a_m}{r^2}-2im\frac{b_m}{r^2}\right)\Biggr]=0,\\
\nonumber\mathcal{L}b_m&-(1+\Gamma\tau-im\Omega\tau)Ga_m\\
\label{b_m}
&-c_\tau\eta\turb \left(\widetilde{\nabla}^2b_m-m^2\frac{b_m}{r^2}+2im\frac{a_m}{r^2}\right)=0,
\end{align}
where 
\[
\mathcal{L}=(1+\Gamma\tau-im\Omega\tau)(\Gamma+im\omtilde).
\]
We note that $\alpha$ is, in general, a function of $r$, $\phi$ and $z$,
whereas $\omega$ and $G$ are assumed to be functions of $r$ alone. 
When obtaining Eq.~\eqref{b_m} from 
Eq.~\eqref{telegraph_phi_corot}, the summation can be dropped because the only $\phi$-dependence 
that occurs in this equation is in the common factor $\exp{(im\phi)}$, whereas $\alpha$ depends on 
$\phi$ in Eq.~(\ref{a_m}).

To make further progress, we model $\alpha$ as an $n$-armed, rigidly rotating global spiral
(thus, in the rotating frame it is time-independent),
\be
\label{alp}
\alpha=\alpha_0[1+\epsilon_\alpha\cos(n\phi-\kappa r)] \equiv
\alpha_0+\alpha_n\Exp{in\phi}+\alpha_{-n}\Exp{-in\phi},
\ee
where
\be
\label{alp_n}
\alpha_n=\tfrac12{\alpha_0\epsilon_\alpha}\Exp{-i\kappa r}, \quad 
\alpha_{-n}=\tfrac12{\alpha_0\epsilon_\alpha}\Exp{i\kappa r}=\alpha_n^*,
\ee
where $\alpha_0$ may depend on $r$ and $z$, 
and asterisk denotes complex conjugate.
With this convention, $\kappa<0$ describes a trailing Archimedean spiral. 
We choose an Archimedean rather than a logarithmic spiral in order
to simplify the calculations; this choice does not affect our conclusions.
The requirement that each coefficient of Eq.~\eqref{a_m} vanishes gives
\[
\begin{split}
\mathcal{L}
a_m
-&c_\tau\eta\turb
\left(\widetilde{\nabla}^2 a_m-m^2\frac{a_m}{r^2}-2im\frac{b_m}{r^2}\right)\\
&=-c_\tau\frac{\del}{\del z}(\alpha_0b_m+\alpha_nb_{m-n}+\alpha_{-n}b_{m+n}).
\end{split}
\]
This implies that components with $m=..., -2n, -n, 0,n, 2n, ...$ are coupled to each other, 
and likewise components with $m=..., -2n+1,-n+1, 1,n+1, 2n+1, ...$, etc.
However, these sets of components are decoupled from one another. 
Components that are coupled to the generally dominant $m=0$ components grow along with it, and are thus 
called `enslaved' components.
One naturally expects the lowest-order enslaved components of order $n$ to dominate the
higher-order enslaved components, as the former are forced directly.
This was in fact borne out in the numerical work of \citetalias{css12}.
Therefore, the $m=\pm n$ components will be the dominant non-axisymmetric 
components for a galaxy with an enslaved $n$-armed spiral pattern. 

Taking $n=2$ (a two-armed $\alpha$-spiral), \citetalias{ms91} and \citet[][hereafter \citetalias{sm93}]{sm93} truncated the series of coupled 
equations by neglecting $\vert m \vert \ge 3$. 
Thus, only the $m=\pm1$ are coupled and likewise $m=0$ is coupled to $m=\pm2$. 
Here we truncate the series in the same way; 
the numerical solutions of \citetalias{css12} confirm that this is an excellent approximation for much of the parameter space (however, see below). 
The case of $m=\pm1$ has already been considered in \citetalias{ms91}.
Thus, we focus on the $m=0,\pm2$ case here, 
for which the governing equations are
\begin{align}
\label{m0a}
(1+\Gam\tau)\Gam a_0-c_\tau\eta\turb
\widetilde{\nabla}^2 a_0
&=-c_\tau\frac{\del}{\del z}(\alp_0b_0+\alp_2b_{-2}+\alp_{-2}b_2),\\
\label{m0b}
(1+\Gam\tau)\Gam b_0-c_\tau\eta\turb
\widetilde{\nabla}^2 b_0
&=(1+\Gam\tau)Ga_0,
\end{align}
while, for $m=\pm2$ and $n=2$,
\begin{align}
\label{m2a}
\nonumber L
a_{\pm2}
-c_\tau\eta\turb&\left(
\widetilde{\nabla}^2 a_{\pm2}-4\frac{a_{\pm2}}{r^2}\mp4i\frac{b_{\pm2}}{r^2}\right)\\
&=-c_\tau\frac{\del}{\del z}(\alp_0b_{\pm2}+\alp_{\pm2}b_0),\\
\label{m2b}
\nonumber L
b_{\pm2}
-c_\tau\eta\turb&\left(
\widetilde{\nabla}^2 b_{\pm2}-4\frac{b_{\pm2}}{r^2}\pm4i\frac{a_{\pm2}}{r^2}\right)\\
&=(1+\Gam\tau\mp2i\Omega\tau)Ga_{\pm2},
\end{align}
where
\[
L=(1+\Gam\tau\mp 2i\Omega\tau)(\Gam\pm2i\omtilde).
\]

\subsubsection{Non-axisymmetric modes under the no-$z$ and WKBJ approximations}
\label{NAMU}\label{sec:noz}
We note that for the modes involving even components $m=0,\pm2$ there are six coupled partial differential equations.
\citetalias{ms91} solved the corresponding simpler problem for the modes involving odd ($m=\pm1$) components, 
which involves four coupled PDEs, 
using WKBJ methods, but they did not do the same for the modes involving even components.
Thus we would like to get analytical insights into these perhaps more important modes. 
Therefore we proceed as follows:
Our plan is to first solve for $a_{\pm2}$ and $b_{\pm2}$. 
We do this in Appendix~\ref{sec:details} by using the no-$z$ approximation for the 
$z$-derivatives (\citetalias{sm93}; \citealt{mo95, phillips01}),
and a WKBJ-type approximation to handle radial 
diffusion.
This yields algebraic equations which can be solved to obtain explicit expressions for $a_{\pm2}$ 
and $b_{\pm2}$.
We then substitute them into the $m=0$ equations, \eqref{m0a} and \eqref{m0b}, 
and apply the no-$z$ approximation to obtain Schr\"{o}dinger-type differential equations in $r$ with the effective potential $V(r)$.
Bound states in this potential correspond to exponentially growing solutions, 
and these are obtained using the WKBJ approximation.
This is done iteratively using numerical methods, so that the method is semi-analytical 
rather than analytical. Thus
we avoid solving the six coupled differential equations \eqref{m0a}--\eqref{m2b}.
We shall see that this procedure, in which four of the six differential 
equations are approximated as algebraic equations, 
works well for the case where the vertical turbulent diffusion is much stronger than the radial 
turbulent diffusion, i.e., for solutions of a large radial scale (wavelength) in a thin disc.

As shown in Appendix~\ref{sec:details}, solutions of Eqs.~(\ref{m2a}) and (\ref{m2b}) can be represented as
\begin{align}\label{apm2}
a_{\pm2}=&-\frac{|D_0|b_0\epsilon_\alp}{2}\sqrt{\frac{\Atilde^2+\Btilde^2}{A^2+B^2}}\exp{[\mp 
i(\kappa r+\beta+\betatilde)]},\\
\label{bpm2}
b_{\pm2}=&\frac{|D_0|b_0\epsilon_\alp}{2}\frac{1}{\sqrt{A^2+B^2}}\exp{[\mp i(\kappa r+\beta)]},
\end{align}
where $A$, $B$, $\Atilde$ and $\Btilde$, as well as $\beta$ and $\betatilde$, are functions of position defined in Appendix~\ref{sec:details}.
Here the dynamo number is defined as
\be
D_0=\frac{\alpha_0Gh^3}{\eta\turb^2}=\frac{\alpha_0Gt\diff^2}{h}<0,
\ee
where $t\diff=h^2/\eta\turb$ is the vertical turbulent diffusion time scale.

Once $a_{\pm2}$ and $b_{\pm2}$ are obtained,
we can solve Eqs.~(\ref{m0a}) and (\ref{m0b}) for $a_0$ and $b_0$. 
Following \citetalias{rss88} and \citetalias{ms91} by invoking the thinness of the disc, 
we factorise the solution into the local ($\widetilde{a}$ and $\widetilde{b}$) and global ($q$) parts,
\be
\label{local-global}
a_0(r)=\widetilde{a}q(r),\qquad b_0(r)=\widetilde{b}q(r).
\ee
Generally, the local solution depends on $z$, i.e., 
$\widetilde{a}(z)$ and $\widetilde{b}(z)$. 
However, the $z$-dependence has been removed using the no-$z$ approximation, 
and the local solution is a vectorial constant.

\subsubsection{The local solution}
We shown in Appendix \ref{sec:Gamma} that the local equations can be written as
\begin{align}
\label{localr_noz}
\left(\gamma+\frac{\pi^2c_\tau}{4t\diff}\right)\atilde
=&-c_\tau\frac{2\alp_0}{\pi h}\btilde\left[1+\frac{|D_0|\eps_\alp^2}{4}(X_2+X_{-2})\right],\\
\label{localphi_noz}
\left(\gamma+\frac{\pi^2c_\tau}{4t\diff}\right)\btilde
=&(1+\Gam\tau)G\atilde,
\end{align}
with $X_{\pm2}$ defined in Appendix \ref{sec:details}.
These homogeneous equations in $\atilde$ and $\btilde$ are easily solved to yield  
$\btilde/\atilde$ and $\gamma$.  From Eq.~\eqref{localphi_noz}, we obtain
\[
\frac{\btilde}{\atilde}=\frac{(1+\Gam\tau)Gt\diff}{\gamma t\diff+\tfrac14 c_\tau\pi^2}.
\]
Then the vanishing of the determinant of the coefficients of 
Eqs.~\eqref{localr_noz} and \eqref{localphi_noz} yields
\begin{equation}
\label{gamma}
\gamma=t\diff^{-1}\left\{-\frac{\pi^2c_\tau}{4}
\pm\sqrt{\frac{2c_\tau}{\pi}(1+\Gam\tau)|D_0|
\left[1+\frac{\epsilon_\alp^2|D_0|A}{2(A^2+B^2)}\right]}\right\},
\end{equation}
where the positive sign in front of the square root provides growing solutions, $\gamma>0$.

With $\eps_\alp=0$ (axisymmetric forcing) and $\gamma=0$ (neutral stability of the mean magnetic
field), we obtain an estimate for the local critical dynamo number, 
i.e., the minimum magnitude of the dynamo number for the field to grow rather than decay,
\[
|D_{0,\crit}|= \frac{\pi^5c_\tau}{32(1+\Gamma\tau)}\approx\frac{9.6c_\tau}{1+\Gamma\tau}.
\]
It is slightly larger than the $D_{0,\crit}=8$, 
obtained by the numerical solution of the $z$-dependent versions of Eqs.~\eqref{localr_noz} and 
\eqref{localphi_noz} for $c_\tau=1$, $\tau=\eps_\alp=0$ \citepalias{rss88}; we note that $\Gamma\tau\ll1$.

For $\eps_\alp=0$, the magnetic pitch angle, given by Eq.~\eqref{pitch}, can be expressed as
\be
\label{pitch_axisym}
\tan p_{B,0}=\frac{\atilde}{\btilde}=-\sqrt{\frac{2c_\tau}{\pi(1+\Gamma\tau)h}\frac{\alpha_0}{|G|}}.
\ee
A similar result can be obtained from perturbation theory \citep{sss07}.
With 
\begin{equation}\label{Kr}
\alpha_0=l^2\omega/h
\end{equation}
and $G=-\omega$ (flat rotation curve), this becomes
\[
\tan p_{B,0}=-\frac{l}{h}\sqrt{\frac{2c_\tau}{\pi(1+\Gam\tau)}}.
\]
The azimuthal average of the pitch angle in the kinematic regime obtained numerically 
in \citetalias{css12}, 
where Eq.~\eqref{Kr} was used, agrees very well with this formula.
This can be seen by using the relevant parameters from that paper in the above equation, 
and then comparing with Fig.~8d of that paper.
On the other hand, if, for example, $\alp_0$ and $h$ are independent of radius, 
then Eq.~\eqref{pitch_axisym} with $|G|=\omega$ implies
that $|p_{B,0}|$ increases with radius, as $\omega$ is a decreasing function of $r$.

\subsubsection{The global solution}
From Appendix~\ref{sec:Gamma}, the global equation is given by
\begin{equation}
\label{global}
(1+\Gam\tau)\Gam q=\gamma q+c_\tau\eta\turb\left[\frac{(rq)'}{r}\right]',
\end{equation}
where prime stands for $d/dr$.
Now, substituting Eq.~\eqref{gamma} into Eq.~\eqref{global}, 
we obtain a Schr\"{o}dinger-type equation for $q(r)$,
\begin{equation}
\label{q}
-\Del^2_rq+V(r)q(r)=Eq(r),
\end{equation}
where $\Del_r^2=(rq')'/r$ is the radial part of the Laplacian in cylindrical coordinates.
Here, the `potential' is given by
\begin{equation}
\label{V}
\begin{split}
V&=\frac{1}{r^2}-\frac{\gamma t\diff}{c_\tau h^2}\\
&=\frac{1}{r^2}
	+\frac{\pi^2}{4h^2}
			-\frac{1}{h^2}
					\sqrt{
					\frac{2(1+\Gam\tau)|D_0|}{\pi c_\tau}
						 \left[1+\frac{\epsilon_\alp^2|D_0|A}{2(A^2+B^2)}\right]},
\end{split}
\end{equation}
and the `energy' eigenvalue, by
\begin{equation}
\label{E}
E=-\frac{(1+\Gam\tau)\Gam t\diff}{c_\tau h^2}.
\end{equation}
Note that bound states in the `potential' $V(r)$ with $E<0$ correspond to growing modes with $\Gamma>0$.
We solve this equation using the WKBJ theory: 
in terms of a scaled variable $x$, introduced via $r=h\f\Exp{x}$ (with $h\f=\const$, 
$-\infty<x<\infty$ and $dx=dr/r$), 
Eq.~(\ref{q}) reduces to
\[
\frac{d^2q}{dx^2}+p(x)q(x)=0,
\] 
where
\[
p(x)=h\f^2\Exp{2x}[E-V(x)].
\]
The boundary conditions are $q=0$ at $r=0$ and $r \to \infty$ (or $x \to \pm \infty$). 
Suppose that $p(x) =0$ for $x=x_{\pm}$ (corresponding to $r_{\pm}$), with $x_{-} < x_{+}$ 
(for our purposes, it is sufficient to consider the case of just two roots). 
The standard WKBJ theory yields the quantization condition
\begin{equation}
\label{quantization}
\displaystyle\int^{r_+}_{r_-}\frac{\sqrt{p(r)}}{r}dr=\displaystyle\int^{r_+}_{r_-}[E-V(r)]^{1/2}dr=\frac{\pi}{2}(2k+1),
\end{equation}
with $k=0,1,2,\ldots$, which can be used to obtain the growth rate $\Gamma$.
Note, however, that $\Gamma$ enters both the `potential' and the `energy',
so the WKBJ approach has to be supplemented with an iteration procedure to converge on the correct 
value of $\Gamma$, in addition to the iterations over $\beta$ and $\betatilde$ discussed in Sect.~\ref{IE} and Appendix~\ref{sec:details}.

As we will see below, the $\phi$-dependent term in the $\alpha$ effect leads to a deepening 
of the potential well near the co-rotation radius, 
allowing strong non-axisymmetric modes that co-rotate with the $\alpha$-spiral to exist there.
Outside this region (i.e., far from the co-rotation radius) axisymmetric modes will 
dominate, while non-axisymmetric modes will be weaker.
The asymptotic solutions developed here apply to the corotating magnetic modes in the kinematic regime, 
but the numerical simulations of \citetalias{css12} are, of course, not restrictive.

Substituting Eqs.~\eqref{apm2} and \eqref{bpm2} into Eq.~\eqref{Bsum}, and using 
Eqs.~\eqref{local-global}, we can write the overall solution as
\begin{align}
\label{mbr}
\mbr&=\atilde q \,\Exp{\Gam t}
\left[1-\frac{\btilde}{\atilde}\sqrt{\frac{\Atilde^2+\Btilde^2}{A^2+B^2}}
\epsilon_\alp|D_0| \cos(\psi-\widetilde{\beta})\right],\\
\label{mbp}
\mbp&=\btilde q \,\Exp{\Gam t}
\left[1+\frac{1}{\sqrt{A^2+B^2}}\epsilon_\alp|D_0|\cos\psi\right],
\end{align}
where
\[
\psi=2\phi-\kappa r-\beta.
\]
Magnetic field lines of the mean field are expected to be trailing spirals since $G<0$, 
which implies $\btilde/\atilde<0$. 
Therefore, the overall phase difference between $\mbr$ and $\mbp$ has magnitude $\betatilde$
[without the minus sign introduced in front of $\mean{a}$ in Eq.~\eqref{ab}, there would have been 
an extra phase difference $\pi$]. The pitch angle of the magnetic field, defined as
\be
\label{pitch}
p_B=\arctan\frac{\mbr}{\mbp},
\ee
is independent of $q$.

Each component of the mean magnetic field consists of an axisymmetric part 
and that having the $m=2$ symmetry,
\[
\mean{B}_i=\mean{B}_i^{(0)}+\mean{B}_i^{(2)}\cos[2(\phi-\phi_i)],
\]
where $i=r,\phi$. The magnitude of the $r$-component of the field is maximum 
where $2\phi-\kappa r-\beta-\betatilde=0$,
while the magnitude of the $\phi$-component is maximum 
for $2\phi-\kappa r-\beta=0$.
On the other hand, the magnitude of $\alpha$ is maximum  
where $2\phi-\kappa r=0$.
Therefore, the ($\pi$-fold degenerate) phase differences between the $r$ and $\phi$ components of 
the magnetic field and the $\alpha$-spiral are given, respectively, by
\[
\Delta_r=\frac{\beta+\betatilde}{2}, \qquad \Delta_\phi=\frac{\beta}{2}.
\]

\subsubsection{Non-enslaved non-axisymmetric modes}
Non-axisymmetric modes that rotate at an angular speed different from that of the 
$\alpha$-spiral and are not enslaved (they
rotate at approximately the local angular velocity 
at the radius where their eigenfunctions are maximum,) 
can also be maintained but they are
sub-dominant everywhere \citepalias{ms91}. 
Such modes have been observed to exist in axisymmetric discs (\citealt{rss88}, \citealt{mo96}, \citetalias{css12}).
In Paper I we showed that the growth of such modes in an axisymmetric disc requires somewhat special parameter values,
and that in any case, such modes decay in the nonlinear regime.
For a non-axisymmetric disc, such modes could be found by replacing $\Gamma$ in Eq.~\eqref{Bsum} 
(and hence in subsequent equations) with $\Gamma\real -im\Gamma\imag$,
where $\Gamma\real$ is the growth rate and $\Gamma\imag$ is the angular velocity 
of the mode in the reference frame that co-rotates with the $\alpha$-spiral.
However, with this more general approach, the WKBJ treatment of Sect.~\ref{sec:Gamma} would, 
strictly speaking, have to be replaced by a \textit{complex} WKBJ treatment.
Given the numerical result that non-corotating non-axisymmetric modes are less important than corotating non-axisymmetric modes,
we have chosen to leave the study of the former for the future.

\subsection{Illustrative example}\label{sec:IE}
\begin{figure*}
  \includegraphics[width=160mm]{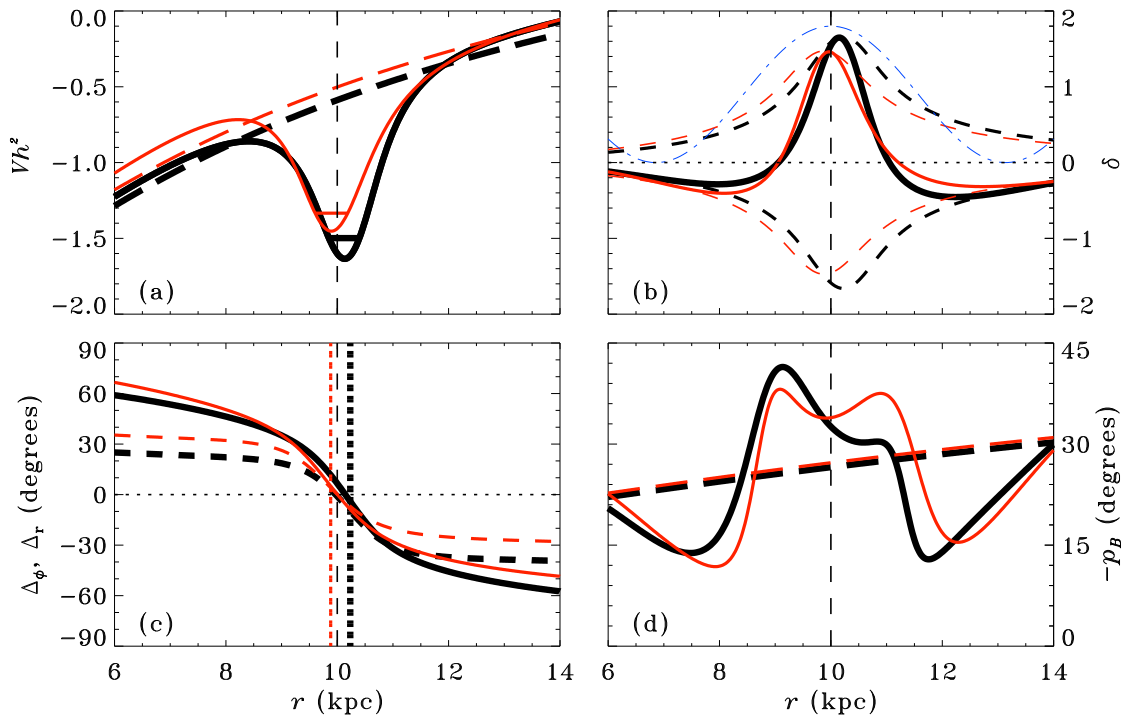}
  \caption{{\bf (a)} The potential \eqref{V} of \eqref{q} for $\tau=2l/u$ (thick, black, solid 
  curve) and $\tau\rightarrow0$ (thin, red, solid curve), near the co-rotation radius (dashed vertical line). 
  The `energy' eigenvalues \eqref{E} of \eqref{q} are represented by horizontal lines inside the wells. 
  The axisymmetric part of the potential is also shown as a dashed line in both cases.
  {\bf (b)} The solid curves represent the ratio of non-axisymmetric to axisymmetric parts of $\mbp$ 
  at the azimuth $\phi=\phi\corot=\kappa r\corot/2$, while the dashed curves show the corresponding 
  magnitudes of the ratios $\pm \mbp^{(2)}/\mbp^{(0)}$.
  The dash-dotted curve shows the variation of $\alpha(r,\phi\corot)$.
  {\bf (c)} The phase differences, $\Delta_\phi$ (solid) and $\Delta_r$ (dashed), between the components 
  of the magnetic spiral arms and the $\alpha$-spiral arms.
  The radius $r\ma$ where the envelope of $\mbp^{(2)}/\mbp^{(0)}$ is maximum is shown as a dotted vertical line of 
  the appropriate colour and thickness.
  {\bf (d)} Pitch angle $p_B$, shown with the opposite sign, at $\phi=\phi\corot=\kappa r\corot/2$ (solid).
  Also shown is $-p_B$ for the purely axisymmetric case (dashed).
  \label{fig:semi-analytic}}
\end{figure*}
\begin{figure*}
  \includegraphics[width=160mm]{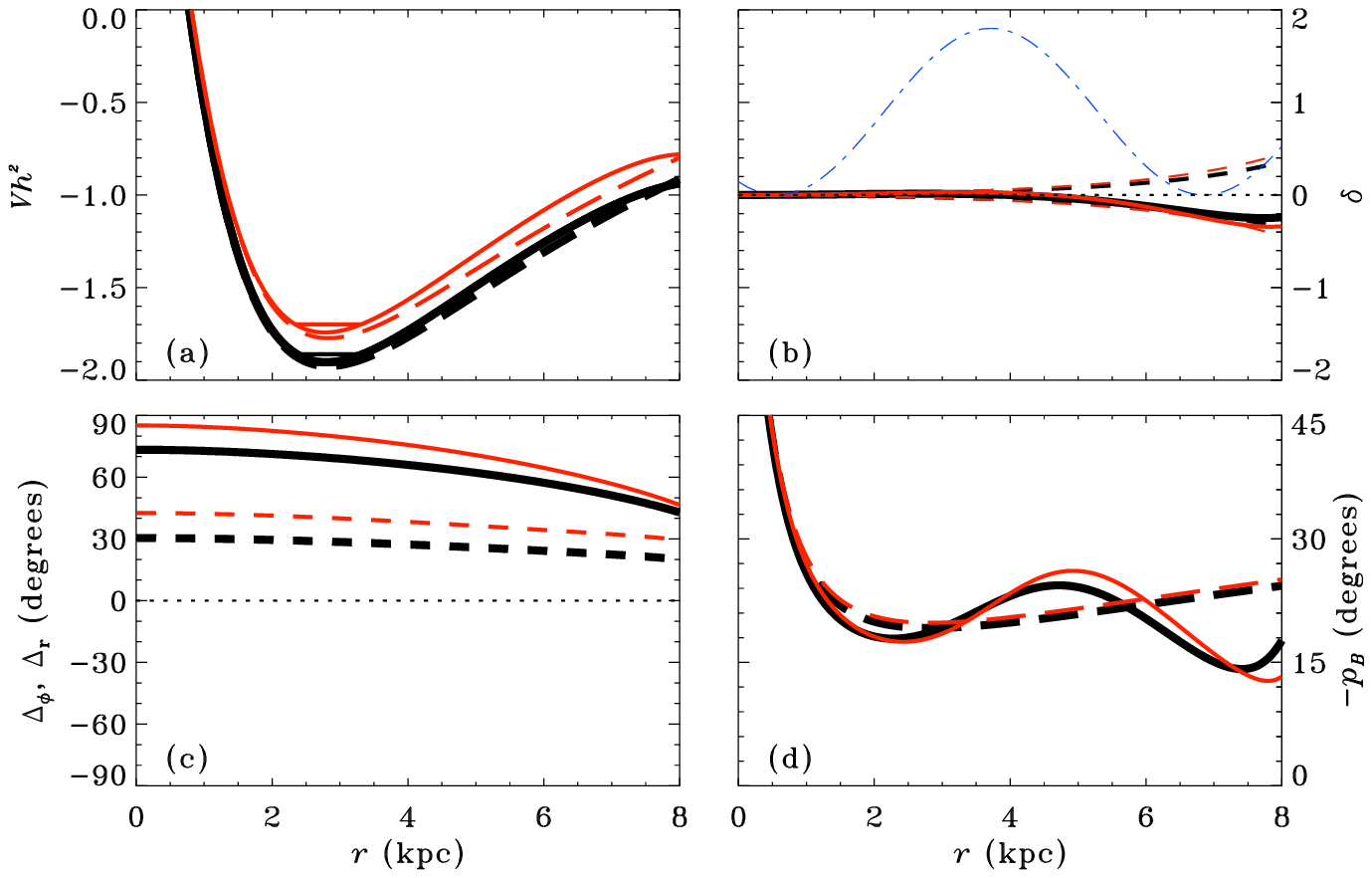}
  \caption{Same as Fig.~\ref{fig:semi-analytic} but 
            for the fastest growing mode in the inner region of the disc, at $r\simeq2.8\kpc$.
  \label{fig:semi-analytic_inner}}
\end{figure*}
We determine $\Gamma$ and the properties of the mean magnetic field by varying $\Gamma$ iteratively until the quantization condition 
\eqref{quantization} is satisfied for $k=0$ (the fastest growing mode).
In addition, there is the complication that $\beta$, $\betatilde$, and hence $\theta_a$, 
$\theta_b$, $A$, $B$, $\Atilde$, $\Btilde$, etc., are not a priori known (see Section~\ref{NAMU} and Appendix~\ref{sec:details}).
We resolve this by iterating these variables as well. 
That is, we start with $\beta'(r)$, $\betatilde'(r)$
and their derivatives equal to zero, 
and at each iteration, we obtain these functions from Eqs.~\eqref{beta} and 
\eqref{betatilde} to use them as input for the next iteration.
The iterations converge if the terms involving $\beta'$ and $\betatilde'$
are small compared with the turbulent diffusion terms that do not vary with $r$.
In the example below, $\beta'$ and $\betatilde'$ are comparable to $\kappa$ near the co-rotation.
However, it can be seen from Eqs.~\eqref{A}, \eqref{B}, \eqref{Atilde} and \eqref{Btilde} 
that the terms involving the radial derivatives are not very important
if the radial diffusion is much weaker than the vertical diffusion, 
that is if $\theta_a'^2,\theta_b'^2\ll\pi^2/4h^2$, which may permit the iterations to converge.
Therefore, in the example below, we choose the disc half-thickness $h$ to be small enough 
(somewhat smaller than the standard value) that the iterations do in fact converge.

In order to determine the potential and growth rates of the corotating 
non-axisymmetric modes, we must first specify the functional forms 
of the scale height $h(r)$, 
angular velocity $\omega(r)$ and $\alpha_0(r)$.
For simplicity, we choose $h=0.1\,{\rm kpc}=\const$.
We also take $\eta\turb=ul/3$, where $u$ and $l$ are the 
rms turbulent velocity and scale,
and adopt $l=h/4=0.025\kpc$, and $u=12ht\diff^{-1}=5\,{\rm km\,s^{-1}}$.
These values are different from those usually adopted, $u=10\,{\rm km\,s^{-1}}$
and  $l=0.1\,{\rm kpc}$. 
Our choice is motivated by the desire to 
improve the convergence of the iterations since our aim here is mainly to illustrate
the solution procedure and to discuss the qualitative properties of the eigen-solutions. 
However, we take care to keep the key parameter, the
vertical turbulent diffusion time, at $t\diff=0.24\Gyr$, 
similar to (about twice smaller than) the standard value.

As in \citetalias{css12}, we use Brandt's rotation curve, 
\[
\omega=\frac{\omega_0}{\left[1+(r/r_\omega)^2\right]^{1/2}},
\]
with $r_{\omega}=20h=2\,{\rm kpc}$, 
and $\omega_0$ set 
to yield 
the circular rotation speed $\mean{U}_\phi=600ht\diff^{-1}=250\,{\rm km\,s^{-1}}$ at 
$r=100h=10\,{\rm kpc}$.
We choose the co-rotation radius to be located at $r\corot=100h=10\,{\rm kpc}$, which corresponds to a pattern speed $\Omega=6t\diff^{-1}=25\,{\rm km\,s^{-1}\,kpc^{-1}}$.

Furthermore, we take $\alpha_0=2.4ht\diff^{-1}=1\,{\rm km\,s^{-1}}$ at all radii.
We also take $\kappa=-0.1h^{-1}=-1\,{\rm kpc^{-1}}$ and $\epsilon_\alp=1$, the latter chosen to 
be large to maximize the strength of the non-axisymmetric modes. We also adopt $c_\tau=1$.

The value of $\tau$ is estimated as $\tau\approx 2l/u\approx 2l^2/3\eta\turb \approx 
2l^2t\diff/3h^2=t\diff/24=9.8\Myr$.
We also consider 
solutions for
$\tau\rightarrow0$, both for reference and also because $\tau$ could be much smaller than 
our estimate of $2l/u$ in some galaxies (though it could be much larger in others).

When describing the geometry of the magnetic modes, it is useful to have in mind that magnetic
lines have the shape of a spiral with the pitch angle defined in Eq.~\eqref{pitch}. These spirals
have to be distinguished from the spirals along which magnetic field strength is maximum. These
can be called magnetic arms or magnetic ridges, and have a different pitch angle 
\citep{brss87,krss89} \citepalias[see also Fig.~VII.8 of][]{rss88}.

The converged results for the modes around co-rotation are shown in 
Fig.~\ref{fig:semi-analytic} for $\tau\rightarrow0$ and 
$\tau=t\diff/24=9.8\Myr$.
(Solutions for intermediate values of $\tau$ are intermediate between the two solutions illustrated.) 
Panel~a shows the potential $V(r)$ of Eq.~\eqref{V}, along with the `energy' 
$-(1+\Gamma\tau)\Gamma t\diff/h^2$ of the fastest growing eigenmode.
Importantly, due to the non-axisymmetric part of $\alpha$, the potential has a local minimum near the 
co-rotation radius. 
Therefore, corotating non-axisymmetric modes can be preferentially excited there.
The finite value of $\tau$ makes the potential deeper and thus enhances the 
growth of non-axisymmetric modes, with a slightly larger growth rate, 
$\Gamma=1.41t\diff^{-1}=5.9\Gyr^{-1}$ for $\tau=t\diff/24$ versus
$\Gamma=1.33t\diff^{-1}=5.6\Gyr^{-1}$ for $\tau\rightarrow0$.

Figure~\ref{fig:semi-analytic}b shows the ratio
\begin{equation}\label{d2}
\delta\equiv\frac{\mbp-\mbp^{(0)}}{\mbp^{(0)}}=\delta^{(2)}=\frac{1}{\sqrt{A^2+B^2}}\epsilon_\alp|D_0|\cos(2\phi-\kappa r-\beta),
\end{equation}
where the equality between $\delta$ and $\delta^{(2)}$ does not hold for numerical solutions, 
which include higher-order azimuthal components,
and where,
\begin{equation}\label{dm}
\delta^{(m)}\equiv\frac{\mbp^{(m)}}{\mbp^{(0)}}\cos\left[m(\phi-\phi_\phi)\right].
\end{equation}
The quantity $\delta^{(2)}$ is given by the second term in the brackets of Eq.~\eqref{mbp}. 
The figures show $\delta$ at the azimuth $\phi=\phi\corot=\kappa r\corot/2$ where one 
of the $\alpha$-spiral arms crosses the co-rotation circle.
The ($\phi$-independent) amplitude (envelope) of this ratio is shown with dashed curves, 
and $\alpha(r,\phi\corot)$ is represented with a dash-dotted curve for reference.
For both finite and vanishing $\tau$, the envelope of $\delta$ exceeds unity near the 
co-rotation,  so that the $|m|=2$ components dominate there.
The dominance is stronger when $\tau$ differs from zero. 
Importantly, the radius $r\ma$ where the envelope of $\delta$ is maximum 
is somewhat larger for the finite $\tau$.

The azimuthal phase differences $\Delta_\phi$ and $\Delta_r$ between the maximum in, 
respectively, the azimuthal and radial magnetic field components, and the maximum of $\alpha$,
are shown in Fig.~\ref{fig:semi-analytic}c. For both $\tau\rightarrow0$ and 
$\tau\ne0$, the magnetic spiral arms (whose phase shift is given approximately by $\Delta_\phi$ since $\mbp$ dominates over $\mbr$) 
cross the $\alpha$-spiral near the co-rotation radius, 
so that each magnetic arm precedes the $\alpha$-arm in azimuth inside the co-rotation circle and lags it at larger radii.
This implies that the magnetic arms are more tightly wound than the material arms.

The $\tau$ effect produces a phase shift in the magnetic arm such that the part of the arm that 
precedes the corresponding $\alpha$-arm is weakened, while the part that lags the $\alpha$-arm is enhanced. 
This can be seen from Fig.~\ref{fig:semi-analytic}c.
For the $\tau\rightarrow0$ case, $r\ma$ (vertical dotted red line) is located slightly inside 
the circles at which $\Delta_\phi=0$ and $\Delta_r=0$.
At these circles, the $\mbp$ and $\mbr$ magnetic arms respectively cross the $\alpha$-arm.
This means that for $\tau\rightarrow0$, the part of the magnetic arm that precedes the 
$\alpha$-arm is 
somewhat
stronger than that which lags.
(For the more realistic numerical solutions of \citetalias{css12}, the preceding and lagging parts 
are of equal strength for $\tau\rightarrow0$ as can be seen in Fig.~8c of that paper.)
On the other hand, for $\tau=2l/u$ in our asymptotic solution, 
we see that $r\ma$ is located to the right of the radii for which $\Delta_\phi=0$ and $\Delta_r=0$,
which tells us that the lagging part of the magnetic arm is stronger than the preceding part.
This is consistent with the findings of \citetalias{css12}.
Thus, the lagging, outer part of the magnetic arm is enhanced for finite $\tau$, while the 
preceding, inner part is weakened.
For example, the phase shift between the magnetic and $\alpha$-arms is 
$\Delta_\phi(r\ma)=5.2^\circ$ for $\tau\rightarrow0$ but $-3.7^\circ$ for $\tau=2l/u$.
However, what we are most interested in is the effect of a finite $\tau$, 
so from now on we will mostly consider the phase {\it difference} between the vanishing and finite $\tau$ cases.
Crucially, we shall show below in Sect.~\ref{sec:PS} that the general behaviour of the phase difference as a function of $\Omega\tau$ 
is nicely reproduced by the asymptotic solution.
Thus, defining 
the difference between the phase shifts obtained for $\tau=0$ and finite $\tau$,
\begin{equation}
\label{Deltadef}
\Delta(\tau)=\Delta_\phi\left[\tau,r\ma(\tau)\right]-\Delta_\phi\left[0,r\ma(0)\right],
\end{equation}
we have $\Delta(2l/u)=-9^\circ$ for the asymptotic solution,
which is of the same order of magnitude as $-\Omega\tau=-14^\circ$.
The phase difference $\Delta(\tau)\simeq-\Omega\tau$, 
obtained here and in the models of \citetalias{css12}, 
is a natural consequence of the finite response time $\tau$ of $\Emf$ to variations in $\alpha$. 

In Fig.~\ref{fig:semi-analytic}d, we plot the pitch angle of the magnetic field $p_B$ at the same 
azimuth $\phi=\phi\corot=\kappa r\corot/2$, 
shown with the opposite sign for presentational convenience.
The pitch angle obtained for the axisymmetric disc (dashed lines) is 
about $p_B=-(20$--$30)^\circ$ in the region shown, and 
increases with radius in agreement with Eq.~\eqref{pitch_axisym}.
Its large magnitude is due to the large, constant value of $\alpha_0$ used in this model, 
whereas the increase of $|p_B|$ with $r$ is due to our choice of a constant disc scale height here, 
as discussed above. 
The azimuthal variation in $\alpha$ leads to a large variation in the pitch angle,
with $-p_B$ being large where $\alpha$ is large and small where $\alpha$ is small (compare with 
the dash-dotted curve for $\alpha$ in Fig.~\ref{fig:semi-analytic}b).
The magnitude of the variation depends on $\epsilon_\alp$.
This variation is explained by the fact that the $\alpha\omega$ dynamo mechanism tends to produce  $|\mbr/\mbp|<1$ due to the velocity shear; 
hence, $|p_B|\propto|\alpha_0/G|^{1/2}$, Eq.~\eqref{pitch_axisym}.
(By contrast, the $\alpha^2$-dynamo would generate rather open magnetic spirals with $|\mbr/\mbp|\simeq1$.)

\subsection{
Modes localised away from the co-rotation radius}\label{sec:modes}
As discussed above and in \citetalias{css12}, 
strong enslaved non-axisymmetric modes are excited near $r=r\corot$.
These modes co-rotate with the $\alpha$-spiral pattern and are enslaved to the $m=0$ component, 
but have very strong $m=\pm2$ components (with the relative amplitude $\delta\sim1$).
Apart from the corotating enslaved non-axisymmetric modes
localised near the co-rotation radius, other types of modes exist within the disc.

The fastest growing modes in the disc are located between $r=0$ and $r=r\corot$, 
near to the radius $r\Dyn$ for which the azimuthally averaged dynamo number is maximum in magnitude.
In this region of the disc, the axisymmetric component dominates, with enslaved non-axisymmetric components
that co-rotate with the $\alpha$-spiral also present, but negligible in comparison to $m=0$.

We illustrate these modes in Fig.~\ref{fig:semi-analytic_inner}, which is similar to 
Fig.~\ref{fig:semi-analytic},
except that it shows the results for the fastest growing mode near $r=r\Dyn$.
Because the potential 
\eqref{V} includes $\Gamma$, the potentials in Figs.~\ref{fig:semi-analytic} and 
\ref{fig:semi-analytic_inner} are different,
though they are qualitatively similar and both have two minima, one near $r\Dyn$ and the other near 
$r\corot$.
Figure~\ref{fig:semi-analytic_inner}a shows that the minimum in the potential is located at a radius 
of $\simeq2.8\kpc$, 
which corresponds almost exactly to the radius $r\Dyn$ for which the magnitude of the dynamo number 
(and hence the shear $G$ since $\alpha$ and $h$ are constant with $r$) is 
maximum. 
The growth rates are $\Gamma=1.73t\diff^{-1}=7.2\Gyr^{-1}$ for $\tau=t\diff/24$ and
slightly smaller, $\Gamma\simeq1.70t\diff^{-1}=7.1\Gyr^{-1}$, for $\tau\rightarrow0$.
The above growth rates are similar to those seen in the more realistic disc models of \citetalias{css12}.
They produce $\sim7$ e-folding times, or a factor $\sim10^3$ in the growth of the field in $1\Gyr$. 
That is, our models can comfortably explain the existence of $\mkG$ strength large-scale fields 
even in young disc galaxies of age $1\Gyr$ if the seed field is of $\nG$ strength.
It can be seen in Fig.~\ref{fig:semi-analytic_inner}b that $\delta\ll1$ around $r\sim r\Dyn$,
which means that non-axisymmetric components are very weak there.
Figure~\ref{fig:semi-analytic_inner}c is shown for completeness, though the large phase shift near 
$r\Dyn$ that it illustrates is of little consequence, 
given that the $m\ne0$ components are so weak there.
Finally, Fig.~\ref{fig:semi-analytic_inner}d illustrates that the azimuthal variation of the pitch 
angle of the magnetic field caused by the $\alpha$-spiral is quite strong, 
even deep inside the co-rotation radius, in agreement with the 
results of \citetalias{css12} (e.g., the bottom row of Figure 10 there).
This can be seen by comparing the solid lines for the non-axisymmetric disc ($\epsilon_\alpha=1$), 
with the dashed lines for the axisymmetric disc ($\epsilon_\alpha=0$).

\section{Comparison with numerical results}
\label{sec:comparison}
\begin{figure}
$
\begin{array}{c}
  \includegraphics[width=0.95\columnwidth]{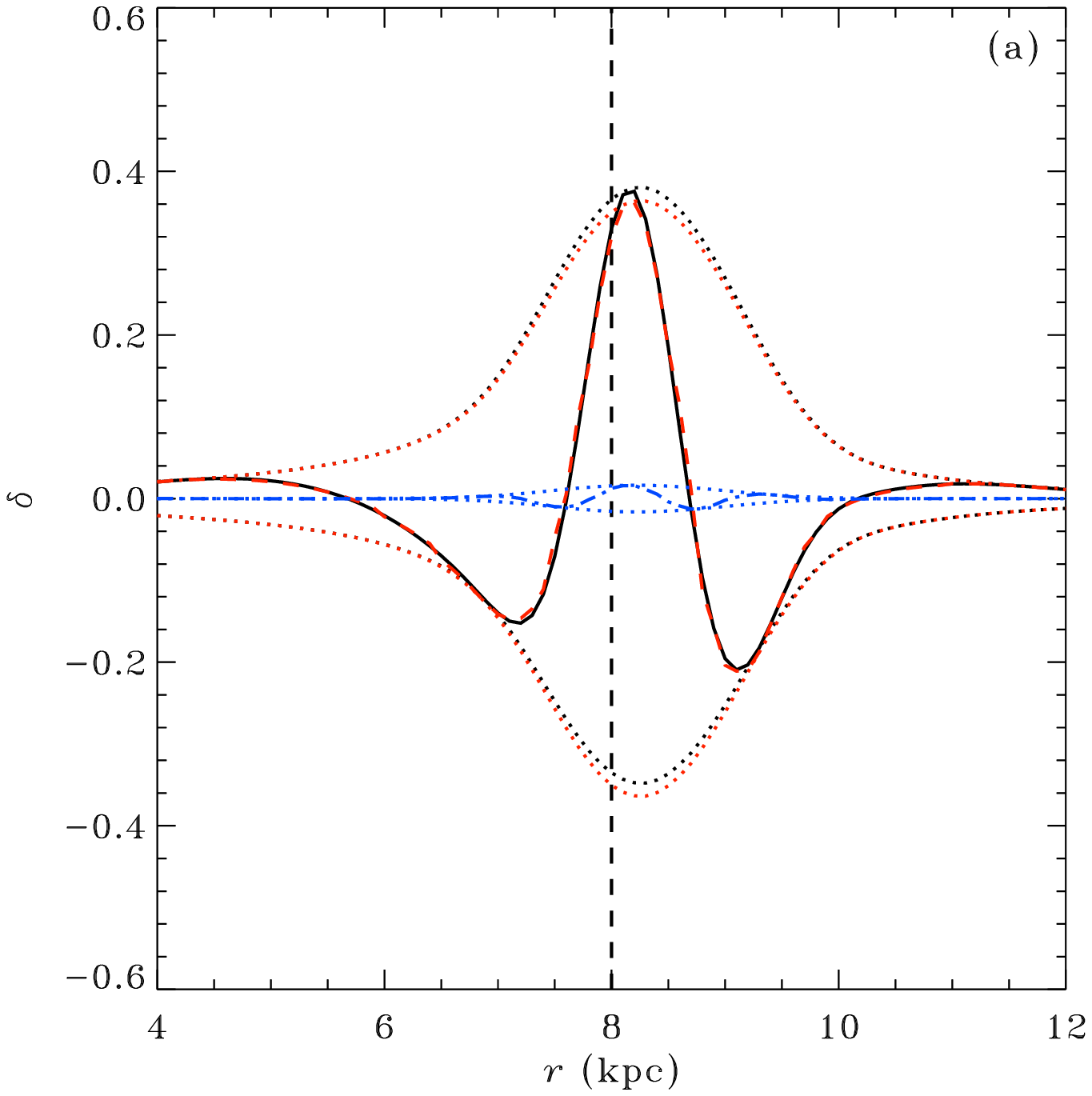}\\
  \includegraphics[width=0.95\columnwidth]{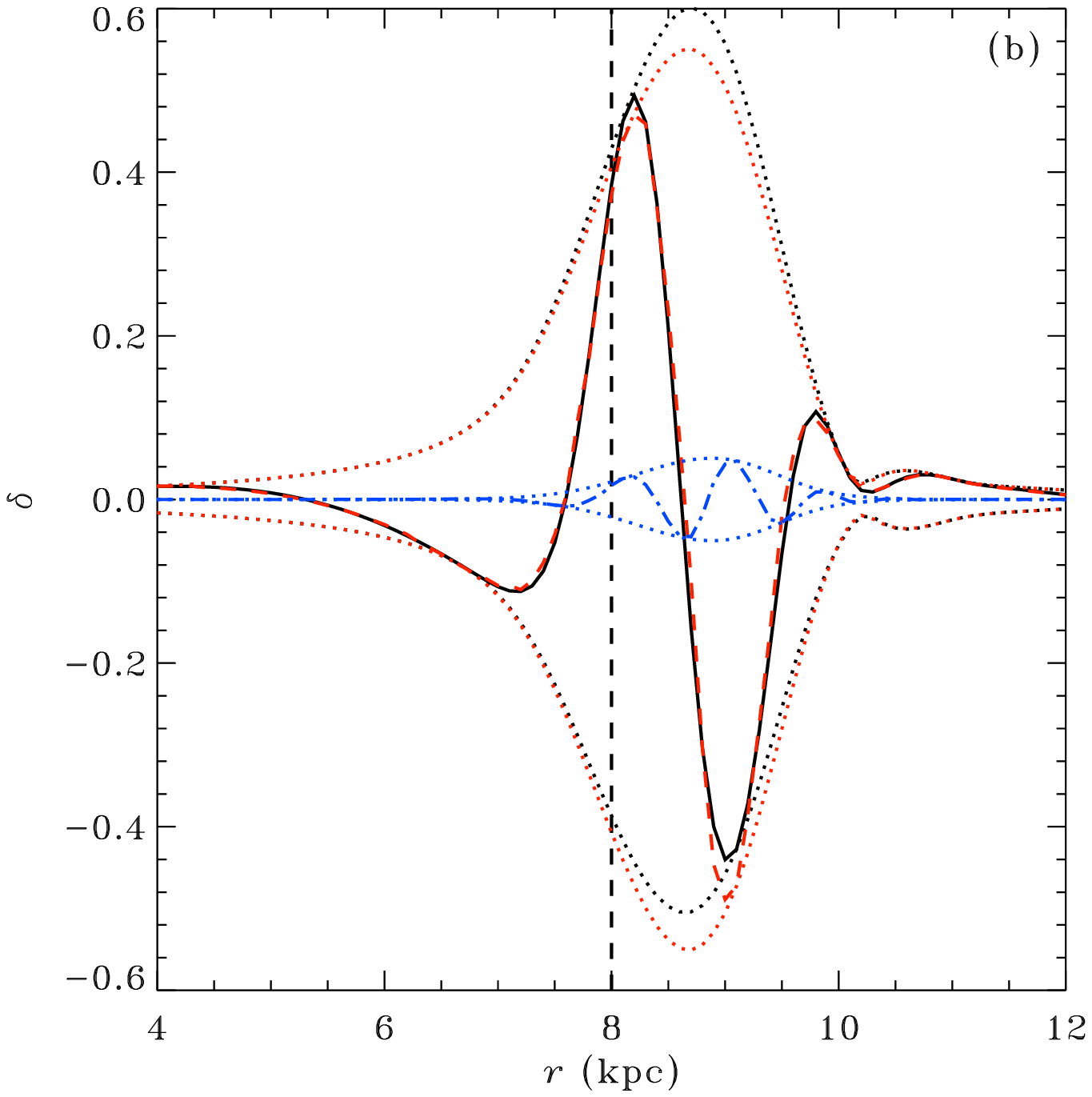}
\end{array}
$
\caption{The relative strength $\delta$ of the non-axisymmetric part of the mean
magnetic field in Model~E of \citetalias{css12}, defined as in Eq.~\eqref{d2}, 
as a function of the galactocentric radius near the 
co-rotation circle at $\phi=\phi\corot$ (black solid), and the envelope of this dependence at
all values of $\phi$ (black dotted). 
The individual contributions of the $|m|=2$ [red dashed for $\delta^{(2)}(r\corot)$ and red dotted for its envelope]
and $|m|=4$ [blue dash-dotted for $\delta^{(4)}(r\corot)$ and blue dotted for its envelope] components are also shown.
{\bf(a)} $\tau\rightarrow0$; {\bf(b)} $\tau=l/u$.
\label{fig:modes}}
\end{figure}
\subsection{Numerical solutions of {P}aper I}
\label{sec:NS}
The semi-analytical solution obtained agrees well, qualitatively, with the numerical solutions of \citetalias{css12},
which can be seen by comparing Fig.~\ref{fig:semi-analytic} with Fig.~8 of \citetalias{css12}, 
which shows the solution of the fiducial non-axisymmetric model of that paper (Model~E) in the kinematic regime.
This is especially true in the $\tau\rightarrow0$ case. 
The numerical solution, however, 
does reveal a greater difference between the finite and vanishing $\tau$ cases than what is seen in the semi-analytical solution.
For example, the phase difference for Model~E of \citetalias{css12}
during the kinematic stage is $\Delta(l/u)=-27^\circ\pm4^\circ$ 
where the range of $\Delta$ reflects the finite resolution of the numerical grid.
This is comparable to but somewhat larger in magnitude, than $-\Omega\tau=-18^\circ$ for that model,
while in our semi-analytical solution, presented above, 
$\Delta$ was found to be comparable, but somewhat smaller in magnitude, than $-\Omega\tau$ for the disc model used.
This apparent discrepancy suggests that there is an additional physical element missing in the asymptotic solution.
In fact, the larger phase shift in the numerical solution can be partly attributed to higher-order even corotating components, especially $m=\pm4$.
Figure~\ref{fig:modes}a shows the
relative strength of the non-axisymmetric magnetic field $\delta$, 
defined in Eq.~\eqref{d2}, at the azimuthal angle $\phi=\phi\corot$,
where the $\alpha$-spiral crosses the co-rotation circle,
for the case $\tau\rightarrow0$ in Model E of \citetalias{css12}.
Fig.~\ref{fig:modes}b shows $\delta(\phi\corot)$ for the case $\tau=l/u$.
These panels are similar to Fig.~8b of 
\citetalias{css12}, with the envelope of $\delta$ over all $\phi$ plotted here as a dotted line.
The quantities $\delta^{(2)}$ and $\delta^{(4)}$ [see Eq.~\eqref{dm}], are also plotted.
It is clear that the $|m|=2$ components dominate, as assumed in the
asymptotic solution. 
However, it is also clear that the ratio of the amplitude of $|m|=4$ to that of $|m|=2$ is larger when $\tau$ is finite.
Moreover, for $\tau=l/u$, the $\delta^{(4)}$ envelope peaks at a larger radius than that of $\delta^{(2)}$.
This suggests that the $|m|=4$ components have the effect of shifting $r\ma$ outward.
Since $\Delta(\tau)$ is sensitive to $r\ma(\tau)$, this can cause a significant increase in $\Delta(\tau)$.

\subsection{
Direct comparison of asymptotic and numerical solutions}
\label{sec:DC}
\begin{figure}
$
\begin{array}{c}
  \includegraphics[width=0.95\columnwidth]{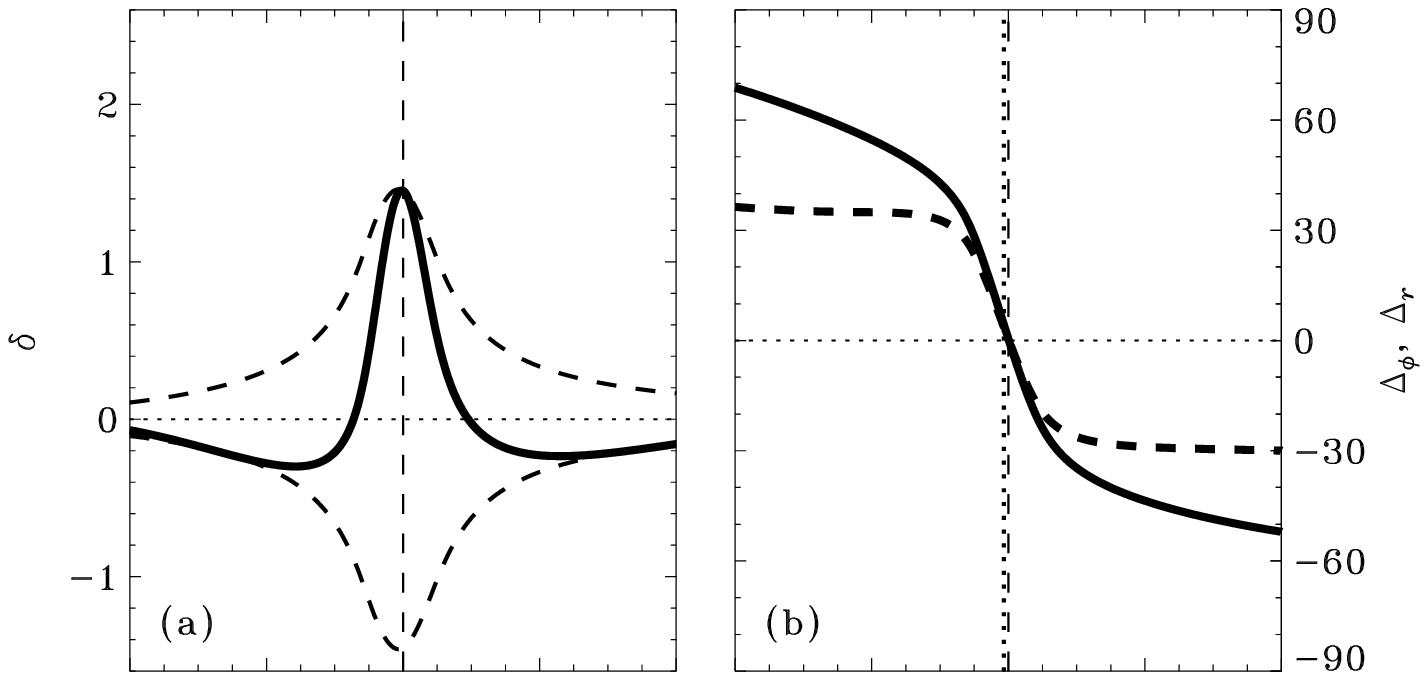}\\
  \includegraphics[width=0.95\columnwidth]{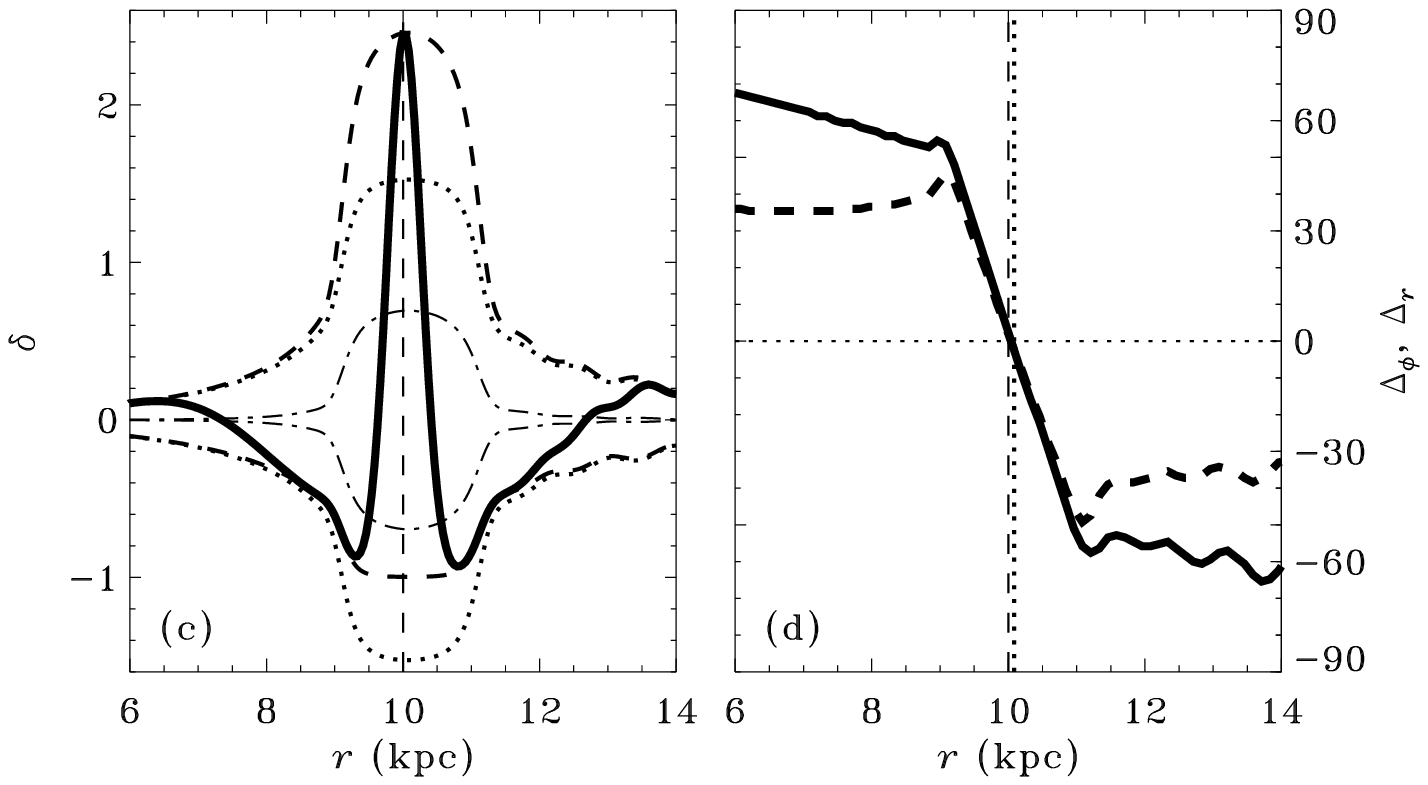}
\end{array}
$
\caption{Direct comparison of asymptotic (a--b) and numerical (c--d) solutions for the same disc model, 
for the case $\tau\rightarrow0$.
Panels a and c are similar to panel b of Fig.~\ref{fig:semi-analytic}, whilst panels b and d are similar to panel c of that figure.
In a and c the $|m|=2$ and $|m|=4$ envelopes are represented by dotted and dash-dotted curves, respectively.
In Panel (d), the curves are smoothed over $0.125\kpc$.
\label{fig:compare_tau0}}
\end{figure}

\begin{figure}
$
\begin{array}{c}
  \includegraphics[width=0.95\columnwidth]{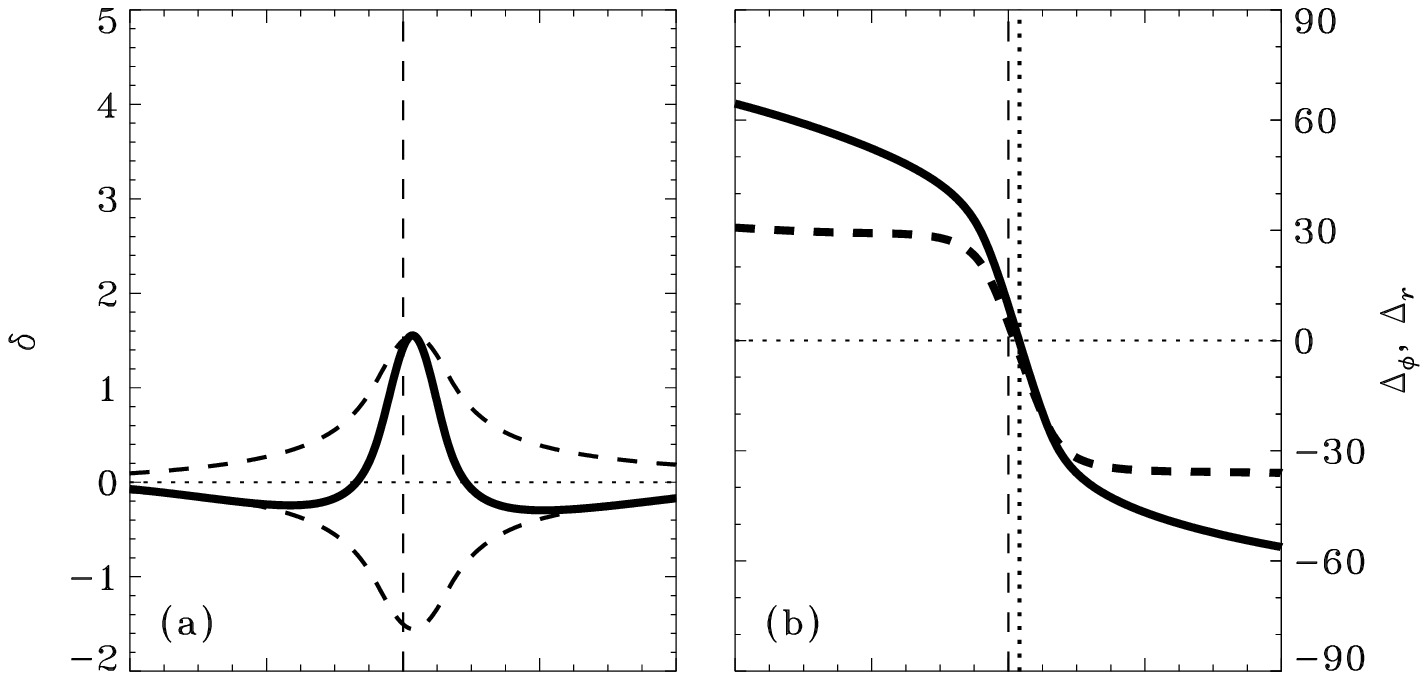}\\
  \includegraphics[width=0.95\columnwidth]{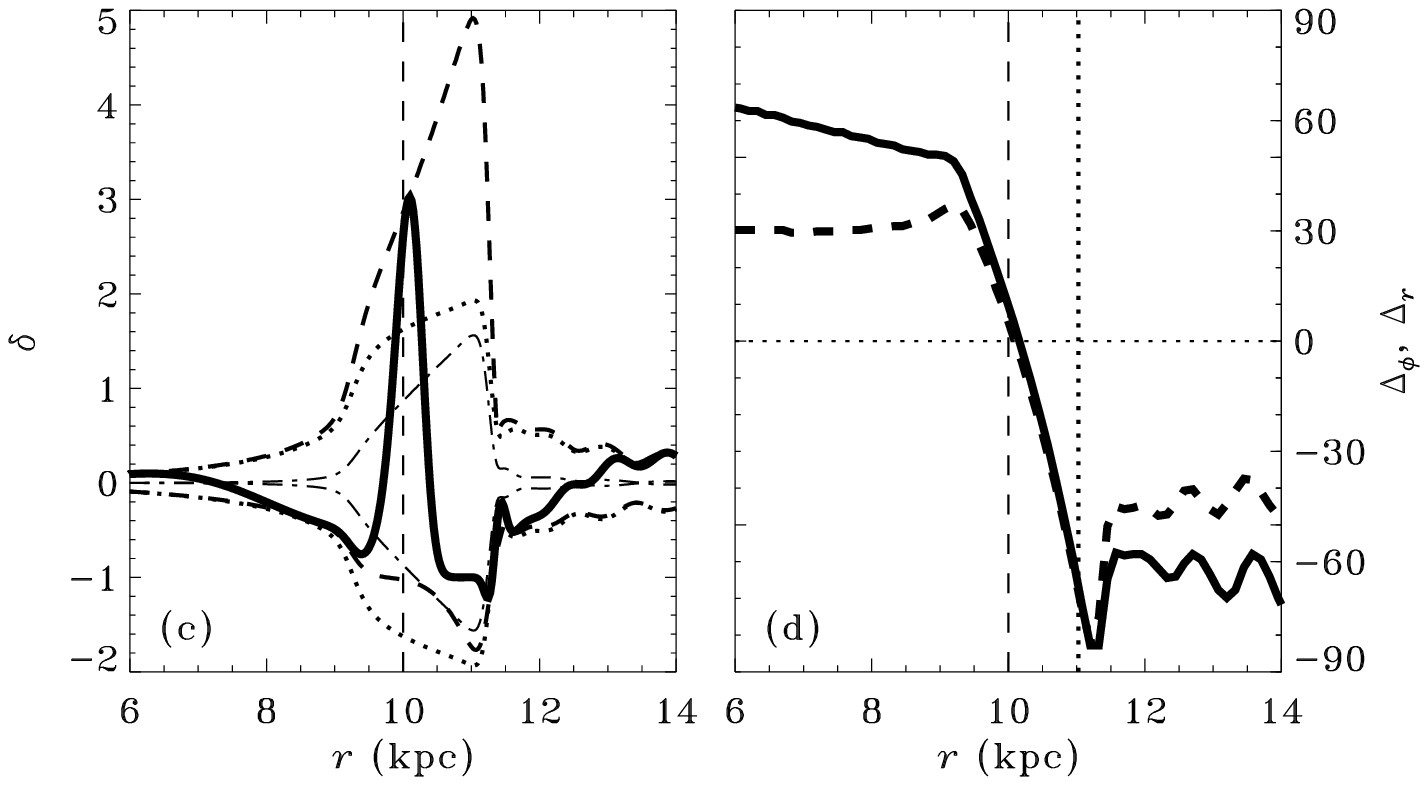}
\end{array}
$
\caption{ 
As in Fig.~\ref{fig:compare_tau0}
but  now for $\tau=l/u$
(note the change in the plotting range of $\delta$),.
\label{fig:compare_taulu}}
\end{figure}

It is also useful to 
compare an asymptotic solution such as that presented above with a numerical solution for \textit{the same disc model} 
(with e.g. a much smaller half-thickness $h$ than the models of \citetalias{css12}).
An example is shown in Figs.~\ref{fig:compare_tau0} and \ref{fig:compare_taulu} for $\tau\rightarrow0$ and $\tau=l/u$, respectively.
The model uses the same parameters as the asymptotic solution above, except that $\epsilon=0.6$ instead of $1$,
for both the asymptotic and numerical solutions.
Plots are shown at $t=10t\diff$, when the relative strength of higher-order \textit{radial} modes has become quite low 
(though they can still be seen as wiggles at $r>11\kpc$).
In Figs.~\ref{fig:compare_tau0}a (asymptotic) and c (numerical), we compare the quantity $\delta$ for the two types of solution, 
while in \ref{fig:compare_tau0}b (asymptotic) and d (numerical), the quantities $\Delta_r$ and $\Delta_\phi$ are compared,
for the $\tau\rightarrow0$ case.
For $\tau\to0$,
the profiles of $\delta$, $\Delta_\phi$ and $\Delta_r$ do not undergo much subsequent evolution.
If, however, $\tau=l/u$,
the envelope of $\delta$ grows and becomes somewhat more asymmetric with time after $t=10t\diff$, 
though the profile of the central (solid) peak at $\phi=\phi\corot$ remains almost unchanged.
Moreover, the asymmetry subsides in the saturated state, and the 
profile of $\delta$
resembles the one shown.

For the case of vanishing $\tau$, the two types of solution are in fairly good agreement.
However, as can be seen by comparing panels a and c of Fig.~\ref{fig:compare_tau0}, 
magnetic arms are more pronounced in the numerical solution than in the asymptotic solution.
This can be attributed to higher order enslaved components in the numerical solution,
as can be seen by comparing the overall envelope (dashed) with the envelopes of the $|m|=2$ (dotted) 
and $|m|=4$ (dash-dotted) components in panel c,

For the $\tau=l/u$ case, shown in Fig.~\ref{fig:compare_taulu}, qualitative features of the solution are in reasonable agreement, 
but the effect of a finite $\tau$ is more dramatic in the numerical solution than in the asymptotic solution.
For instance, $r\ma$ is much larger in the numerical solution due to the asymmetric envelope of $\delta$, resulting in a larger phase shift.
This is seen by comparing panels b and d, 
where it is evident that the dotted vertical line crosses the solid and dashed lines 
at much larger values of $|\Delta_\phi|$ and $|\Delta_r|$ in the numerical solution.
As with the solutions from \citetalias{css12}, enslaved components with $|m|>2$ are stronger in the finite $\tau$ case than in the $\tau\rightarrow0$ case,
as can be seen by comparing the dash-dotted envelopes for the $|m|=4$ components in Figs.~\ref{fig:compare_taulu}c and \ref{fig:compare_tau0}c.
The assertion, made above, that higher-order enslaved components are responsible for enhancing the phase shift is therefore strengthened.

The fact that non-axisymmetric components can be more easily generated in a thinner disc, 
where the difference in the global mode structure is less important because of the stronger dominance
of the local dynamo action (due to the shorter magnetic diffusion time across the disc),
is well known \citepalias[Sect.~VII.8 in][]{rss88}. 
However, the importance of the finite relaxation time in this respect was not appreciated earlier.
We emphasize that this disc model is used for the sake of illustration only, and is not as realistic as the models of \citetalias{css12},
where enslaved components with $|m|>2$ are anyway much less important.

It is also interesting to compare the growth rates of the asymptotic and numerical solutions. 
Both types of solution give the same growth rate for the inner mode located near $r\Dyn$: 
$\Gamma=1.73t\diff^{-1}$ for $\tau=l/u$ and $\Gamma=1.72t\diff^{-1}$ for $\tau\rightarrow0$.
The magnetic field in the inner disc is maximum
at $r=2.9\kpc$ in the numerical solution, for both  values of $\tau$ considered,
in close agreement with the asymptotic solution where the minimum of the potential is situated at $2.8\kpc$ in both cases.
For the modes near the co-rotation radius, the semi-analytical model predicts the growth rates 
of $\Gamma=0.99t\diff^{-1}$ for $\tau=l/u$ and $\Gamma=0.97t\diff^{-1}$ for $\tau\rightarrow0$, 
while the numerical solution gives $\Gamma=1.06t\diff^{-1}$ and $\Gamma=1.02t\diff^{-1}$, respectively.
Again, the agreement between the prediction of the asymptotic solution and the numerical solution is quite satisfactory.
The growth rates for non-axisymmetric modes near the corotation radius are thus about 
half as large as those of the inner axisymmetric modes.

Although one could, in principle, include 
magnetic components with $|m|>2$ in the asymptotic solution presented above, 
including $m=\pm4$ adds four new differential equations as well as new terms to the 
existing equations \eqref{m2a}, increasing the complexity of the model.
Moreover, from our analytical expressions \eqref{apm2} and \eqref{bpm2} for $a_{\pm2}$ and $b_{\pm2}$ 
we find that the WKBJ-type approximations \eqref{tightwinding} turn out to be strictly valid 
only out to $\sim0.5\kpc$ on either side of the minimum (near $r=r\corot$) in the potential.
At those locations, the magnitude of the rate of change of the amplitude with $r$ is comparable to 
that of the phase, in violation of Eq.~\eqref{tightwinding}. This limitation becomes more severe for $|m|\geq4$.
We take the view that the asymptotic solution presented above anyway does a very good job 
of reproducing  the key qualitative features of the numerical solution, and thus we do not include the $m=\pm4$ components.

\subsection{Competition between the inner and outer modes}
Due to their larger growth rate, $m\ne0$ components in the inner disc, 
although weaker than the $m=0$ component there,
soon come to be stronger than the $m\ne0$ components near the co-rotation circle 
(assuming a relatively uniform seed field).
In fact, in the kinematic regime the local extremum of
the magnetic field near the co-rotation radius is gradually overcome
by the tail of the dominant $m=0$ eigenfunction (which has it maximum near $r\Dyn$).
However, the non-axisymmetric magnetic structure near the
co-rotation radius becomes prominent again in the saturated state
\citepalias[see also Sect.~5 of ][]{css12}.\footnote{Due to the 
different disc parameters used, 
the $m=0$ eigenfunction becomes dominant
much earlier in the standard Model~E of \citetalias{css12} 
[about $5t\f$ after the simulation is begun, where $t\f$ is defined in \citetalias{css12}] 
than it does in the models presented in this paper (about $70t\diff$ after the simulation is begun).}

\begin{figure}
  \includegraphics[width=0.95\columnwidth]{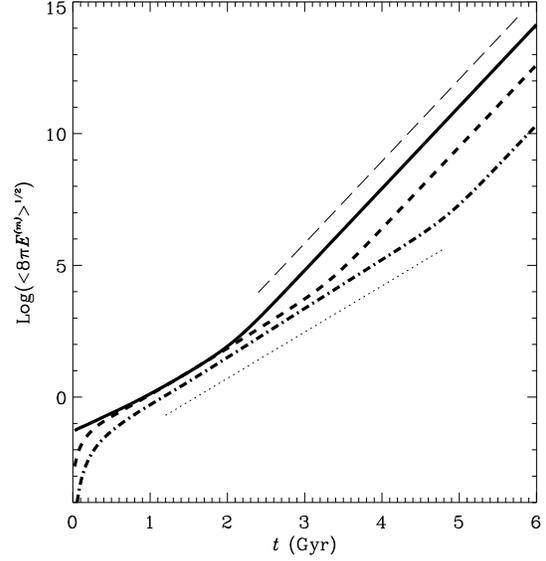}
  \caption{Evolution of the magnetic field strength in each Fourier component for the $\tau\rightarrow0$ case: 
           $m=0$ (solid), $|m|=2$ (short-dashed) and $|m|=4$ (dash-dotted).
           Also plotted are reference lines corresponding to growth rates predicted by the semi-analytical model: 
           $\Gamma=1.72t\diff^{-1}$ (long-dashed) and $\Gamma=0.97t\diff^{-1}$ (dotted).
  \label{fig:Gamma}}
\end{figure}

In Fig.~\ref{fig:Gamma}, the square root of the magnetic energy density
in a component with a given azimuthal symmetry $m$, 
averaged over the area of the disc, is plotted as a function of time for the numerical solution 
(only the $\tau\rightarrow0$ case is shown to avoid clutter, 
though the behaviour is very similar when $\tau$ is finite).
Solid, short-dashed and dash-dotted lines show $m=0$, $|m|=2$ and $|m|=4$ components, respectively.
The long-dashed reference line has slope corresponding to a growth rate of $1.72t\diff^{-1}$, 
while the dotted reference line corresponds to $0.97t\diff^{-1}$.
These reference lines correspond to the growth rates obtained from the asymptotic solution; 
they illustrate the general agreement between asymptotic and numerical solutions discussed above.
Clearly, the modes localised near the co-rotation radius dominate at early times, 
until the modes spreading from $r\Dyn$ catch up at about $t=2\Gyr$.
Under normal circumstances, the inner nearly-axisymmetric mode 
would dominate its counterpart situated near co-rotation right from $t=0$, 
while the $|m|=2$ and $|m|=4$ inner-mode components would quickly come to dominate over their counterparts situated near co-rotation.
We have delayed this inevitable outcome for illustrative purposes using a seed magnetic field 
that is about 1000 times stronger within an annulus of width $4\kpc$ centered on $r=r\corot$.
Thus, the growth rates of both types of mode can be easily calculated from the same graph.
Moreover, it is clear from the figure that, for the modes situated near the co-rotation,
the $|m|=2$ component has almost the same strength as $m=0$, while $|m|=4$ is somewhat weaker (about $0.4$ times as strong). 
For the modes situated near $r=r\Dyn$, on the other hand, $m=2$ is more than $30$ times weaker than $m=0$,
and $|m|=4$ is almost $200$ times weaker than $|m|=2$. 

\begin{figure}
  \includegraphics[width=0.95\columnwidth]{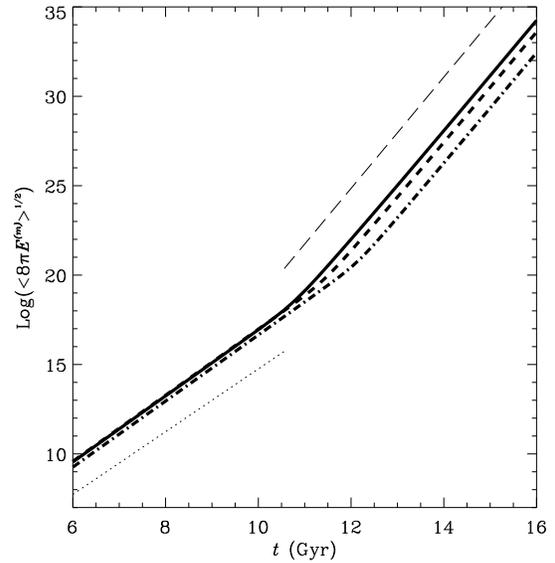}
  \caption{
  As in Fig.~\ref{fig:Gamma} but now with the average taken over the annulus of 
  $4\kpc$ 
  in width
  centred on $r=r\corot$.
  \label{fig:Gamma_rc}}
\end{figure}

In Fig.~\ref{fig:Gamma_rc}, we show a similar plot (obtained without any enhancement of the seed magnetic field near the co-rotation),
but here the averaging is taken over the $4\kpc$-wide annulus around the co-rotation radius.
Clearly, the fastest growing mode localized near $r=r\corot$ dominates until $t\approx12\Gyr$, 
when the tail of the dominant eigenfunction (which peaks near $r=r\Dyn$) overtakes it.
However, as in the solutions of \citetalias{css12}, 
an enhancement in the axisymmetric and non-axisymmetric components of the field near $r=r\corot$ re-establishes itself 
in the saturated state.

In summary, good agreement is obtained between asymptotic and numerical solutions, especially in the growth rates and various qualitative features.
The semi-analytical model presented thus generally succeeds in capturing the key properties of the system.
However, certain details of the solution, especially for finite $\tau$, 
such as the extent of the phase shift between the magnetic and material arms,
are underestimated by the asymptotic analysis.

\begin{figure}
\includegraphics[width=0.99\columnwidth]{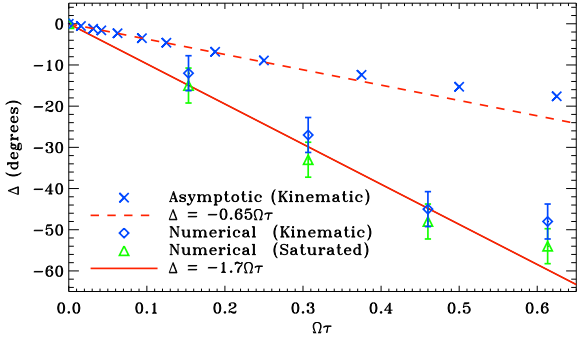}
\caption{
The phase difference $\Delta$ between the solutions with $\tau\neq0$ and $\tau=0$
as a function of $\Omega\tau$. 
Results for the parameter values used in the illustrative example 
are shown with blue crosses; those from
Model~E of \citetalias{css12} 
are shown during the kinematic regime
(at $t=3t_{\rm d,0}$ after the simulation has begun)
and in the steady state.
\label{fig:phaseshift}
}
\end{figure}

\subsection{The phase difference caused by a finite dynamo relaxation time}\label{sec:PS}

It is also interesting to explore how the phase difference $\Delta$ varies with the dimensionless quantity $\Omega\tau$.
Here we retain the disc model (of the asymptotic example or of Model E from \citetalias{css12}) but vary $\tau$.
Figure~\ref{fig:phaseshift} shows the phase difference $\Delta$ as a function
of $\Omega\tau$ for the asymptotic solution of Sect.~\ref{sec:IE}, shown by crosses in the plot,
confirming that $\Delta\approx-C\Omega\tau$ with $C$ a constant of order unity.
It can be seen that the relation is not strictly linear and tends to flatten as $\Omega\tau$ increases, 
but is approximately linear for $\Omega\tau<0.5$, which is the region of parameter space that normally applies to disc galaxies.
Figure~\ref{fig:phaseshift} also presents results from the numerical 
solution of Model~E from \citetalias{css12}, discussed in Sect.~\ref{sec:NS} above,
where the galaxy model is more realistic than in the analytical solution obtained in Sect.~\ref{sec:IE}.
For Model~E, the scaling relation $\Delta\propto\Omega\tau$ (for small $\Omega\tau$) quickly establishes itself in the kinematic regime, 
as soon as the leading eigenfunction becomes dominant. 
Remarkably, it applies to both kinematic dynamo solutions and to the steady state, with more or less the same proportionality constant.
The magnitude of the proportionality constant $C\approx1.7$ (represented by the solid line in the figure) is still of order 
unity, but is significantly larger than 
$C\approx0.65$ (dashed line) found in the asymptotic solution.
Although this difference may be partly attributable to the different disc models used,
the larger value of $C$ in the numerical model is also partly due to higher-order even modes,
as discussed in Sect.~\ref{sec:NS} above.

\subsection{Extrapolation to the nonlinear regime}\label{sec:nonlinear}
\begin{figure}
  \includegraphics[width=1.00\columnwidth]{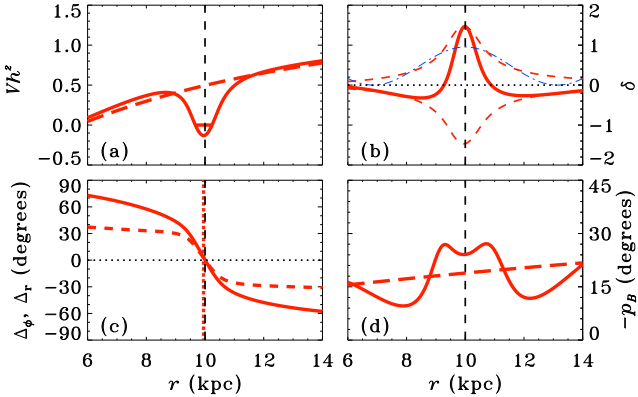}
\caption{
	As in
  Fig.~\ref{fig:semi-analytic} (with $\tau\rightarrow0$) but 
            for $\alpha\f=0.441\kms$,
            which ensures that $\Gamma=0$.
  \label{fig:semi-analytic_marginal}
}
\end{figure}
\begin{figure}
  \includegraphics[width=1.00\columnwidth]{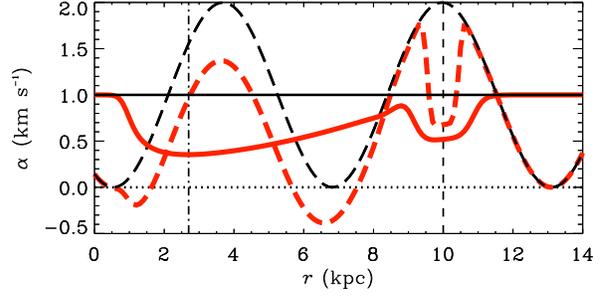}
  \caption{
  The steady-state numerical solution
  based on dynamical quenching, \protect\eqref{akam},  for the same disc 
  parameters as in the asymptotic model. 
  Dashed lines show $\alpha\kin$ (thin black) and
  the total $\alpha$ (thick red) as functions of $r$ 
  at $\phi=\phi\corot$, whereas solid lines are for the respective azimuthal averages.
  The vertical lines indicate the co-rotation radius $r\corot=10\kpc$
  and the radius (about $2.6\kpc$) where magnetic field strength is maximum.
  \label{fig:numerical_saturated}
}
\end{figure}

The asymptotic solution obtained here is strictly valid only in the kinematic (linear) regime. 
However, we find that the general properties of the nonlinear numerical solution are very similar 
to the asymptotic one, and we discuss here the generalisation of the asymptotic solution to nonlinear, saturated states.

In the dynamic nonlinearity model
\citep{pouquetetal76, kleeorinruzmaikin82, gruzinovdiamond94, blackmanfield00, radleretal03, bs05a},
the modification of the $\alpha$-effect 
by the Lorentz force is represented as an additive magnetic contribution $\alpha\magn$, 
so that the total $\alpha$-coefficient becomes
\begin{equation}\label{akam}
\alpha=\alpha\kin+\alpha\magn,
\end{equation}
where $\alpha\kin$ is proportional to the kinetic helicity 
$\overline{\bm{u}\cdot\nabla\times\bm{u}}$ of the random flow and 
is independent of magnetic field, whereas $\alpha\magn$ is proportional to 
the small-scale current helicity $\overline{\bm{b}\cdot\nabla\times\bm{b}}$. 
At an early stage, magnetic force is negligible and $\alpha\approx\alpha\kin$, but $\alpha\magn$ 
(whose sign is opposite to that of $\alpha\kin$) builds up as the mean magnetic field grows. This reduces (quenches) 
the net $\alpha$-effect and leads to a steady state.

As an approximation to the steady-state nonlinear solution, we consider the 
marginal asymptotic solution, i.e., the one with $\Gamma=0$. To obtain 
it, we iterate the values of $\alpha\f$, defined in \eqref{alp}, until 
the quantization condition \eqref{quantization} is satisfied. (We recall
that the procedure to find other solutions is to iterate $\Gamma$.) 
The value of $\alpha\f$ after the iterations
is smaller than the starting one; the difference is attributed to $\alpha\magn$.

Results for $\tau\rightarrow0$ are shown in Fig.~\ref{fig:semi-analytic_marginal},
in a form similar to Fig.~\ref{fig:semi-analytic}.
Equation \eqref{quantization} (with $k=0$) is satisfied for $\alpha\f=0.44\kms$, 
with the starting value $\alpha\f=1\kms$.
The solution is not very different from the kinematic solution of 
Fig.~\ref{fig:semi-analytic},
with somewhat shallower potential well and smaller magnitude of the pitch angle near $r=r\corot$.
The marginal solution for the dominant inner mode near $r=r\Dyn$ obtained in this way has
$\alpha\f=0.35\kms$.

We also solved the mean-field equations with the nonlinearity \eqref{akam} and the same disc parameters
as those used in the marginal asymptotic solution
\citepalias[for details see][]{css12}. 
We include a diffusive magnetic helicity flux with a diffusivity $0.3\eta\turb$ \citep{mitraetal10,candelaresietal11}.
The resulting steady-state form of $\alpha(r)$ is shown in Fig.~\ref{fig:numerical_saturated}.
At the radius where magnetic field has maximum strength,
the azimuthally averaged value of $\alpha$ obtained numerically is $\alpha=0.35\kms$, 
in surprisingly good agreement with that from the asymptotic solution.
At $r=r\corot$, the azimuthal average is $\alpha=0.52\kms$ as compared to $0.44\kms$ 
from the asymptotic solution.
Thus, the steady-state nonlinear state of the galactic mean-field dynamo,
obtained with the dynamical nonlinearity, is reasonably well approximated
by the marginal kinematic solution.

In summary, asymptotic solutions obtained here clarify numerical results of \citetalias{css12} 
and help us to isolate their generic features.
Figure~\ref{fig:semi-analytic} can be compared with Fig.~8 of \citetalias{css12}, 
which shows similar kinematic results, confirming the presence of a strong $|m|=2$ 
magnetic component near $r=r\corot$. 
Around this radius we see clearly from the asymptotic analysis that the effective potential $V(r)$ has a new minimum,
allowing new bound states (or growing modes) of $q(r)$ near co-rotation.
Both asymptotic and numerical solutions reveal stationary modes with even azimuthal components in the frame corotating 
with the spiral pattern of the $\alpha$-coefficient. 
Both have magnetic spiral arms more tightly wound than the $\alpha$-arms, 
with the two types of arm crossing near the co-rotation circle.
Moreover, both numerical and asymptotic results, the latter, e.g., in Eq.~\eqref{gamma}, 
also agree in that the response of the mean magnetic field to the $\alpha$-spiral pattern is 
stronger when $\tau$ is finite. 
Furthermore, they agree in that magnetic arms undergo a 
(against the direction of the galactic rotation) phase shift of order 
$-\Omega\tau$ such that the part of the magnetic arm which trails the $\alpha$-arm is enhanced 
when $\tau$ is finite.
Enslaved components with $|m|>2$, though weak, 
can substantially enhance the phase shift because they are located further out in radius than the $|m|=2$ component,
where magnetic arms lag $\alpha$-arms.

\section{Conclusions}
\label{sec:conclusion}
We have extended non-axisymmetric mean-field dynamo theory in two 
ways.
Firstly, we have obtained semi-analytic solutions for the axisymmetric and the dominant
enslaved non-axisymmetric components 
($m=0$ and $m=\pm2$ for a two-armed $\alpha$-spiral). 
Previously, the only analytical treatment that existed was for the sub-dominant non-enslaved modes (with $m=\pm1$);
Secondly, we have included and explored the effects of the finite relaxation time $\tau$ of the mean 
electromotive force, related to the finite correlation time of the random flow.

We assume that galactic spiral arms lead to the $\alpha$-effect enhanced along spirals 
which may coincide with the material arms or may be located in between them.
The asymptotic solutions employ the WKBJ approximation
applicable to tightly wound magnetic spirals; to obtain explicit expressions for the 
growth rate of the mean magnetic field and its radial and azimuthal distributions we use the
no-$z$ approximation 
to solve the local dynamo equations. 

We find overall good qualitative agreement with the numerical results of \citetalias{css12} for a global, 
rigidly rotating spiral pattern (which is the type of spiral forcing most amenable to analytical treatment)
to grow around the co-rotation radius.
In particular, the asymptotic solution obtained here agrees with the numerical solution
of \citetalias{css12} in the following key features.
\begin{enumerate}
\item[(i)] Strong magnetic modes with the azimuthal wave numbers $m=\pm n$ are supported by the dynamo action, 
which co-rotate with the $\alpha$-spiral (here $n$ is the number of $\alpha$-spiral arms). 
Non-axisymmetric modes are confined to a radial range of a few kiloparsecs around the co-rotation radius 
and enslaved non-axisymmetric components are comparable in strength with the $m=0$ component. 
\item[(ii)] Magnetic arms (understood as ridges of the mean magnetic field strength
that have a spiral shape) cross the $\alpha$-spiral near the co-rotation radius and are more tightly wound than the $\alpha$-arms. 
\item[(iii)] 
In the model considered here, the only effect of the galactic spiral pattern is the
enhancement of the dynamo $\alpha$-coefficient by the spiral pattern. As a result, the 
magnetic lines of the mean field are less tightly wound (larger magnitude of the pitch angle $p_B$)
within the $\alpha$-spirals than between them.
\item[(iv)] The amplitude of the 
non-axisymmetric modes increases with the magnitude of the relaxation time $\tau$.
\item[(v)] Another effect of the finite 
magnitude of $\tau$ is that the part of the magnetic arm that lags behind the $\alpha$-arm with 
respect to the galactic rotation (i.e., outside the co-rotation radius)
becomes stronger than the part that precedes the $\alpha$-arm (inside the 
co-rotation radius). This leads to the overall appearance of magnetic arms that are phase-shifted 
from the $\alpha$-arms (material arms) 
in the direction opposite to the galactic rotation.
\item[(vi)] For $\Omega\tau\la0.5$, the phase shift produced by a finite value of $\tau$ (relative to that for $\tau\rightarrow0$),
is directly proportional to $-\Omega\tau$, with a proportionality constant of order unity.
\end{enumerate}

\section{Discussion for Papers I and II}\label{disc}
Results obtained in \citet{css12} (\citetalias{css12}) and here (Paper~II) 
for a rigidly rotating $\alpha$-spiral
demonstrate the ability of the mean-field dynamo mechanism to produce magnetic arms 
that do not overlap with the material arms in the regions where the former are best pronounced, 
which is outside the co-rotation circle in our model. 
Due to a finite magnitude of the relaxation time of the mean electromotive force, 
$\tau$, the lagging part of the magnetic arm is better pronounced than that ahead of the material 
arm, which can make it difficult to detect their intersection in the observations.
As a result of advection of the mean magnetic field by the differentially rotating gas, 
magnetic arms are, in these models, more tightly wound than the $\alpha$-spiral which drives them,
and are localized to within a few kiloparsecs of the co-rotation circle.
The phase shift between the magnetic and $\alpha$-arms in our model vanishes
at or near the co-rotation radius and then increases, 
first rapidly and then at a lower rate, 
up to a value as large as $60^\circ$ at a distance of a few kpc from the co-rotation circle.

The morphology of magnetic arms in spiral galaxies and their position relative to the 
material arms is rather diverse. One extreme is the galaxy NGC~6946 where magnetic arms appear to 
have the same pitch angle as the optical arms, and the two patterns do not intersect. 
The azimuthal phase shifts between the individual optical arms and their magnetic counterparts 
identified by \citet{fricketal00} are $-26^\circ\pm12^\circ$, $-36^\circ\pm11^\circ$ and 
$-45^\circ\pm17^\circ$ at $r\approx4\kpc$ and, for an outer arm, $-68^\circ\pm17^\circ$ at 
$r\approx8\kpc$,
where negative phase shift corresponds to a position behind a gaseous arm 
\citep[all distances have been reduced to a distance of $5.5\Mpc$ to NGC~6946 -- ][]{Ketal03}.
According to the H$\alpha$ observations and analysis of \citet{FTTBHDCZ07}, the co-rotation 
radius of the outer spiral pattern in this galaxy 
is at about $8.3\kpc$, whereas the interlaced magnetic and optical spiral arms occur
at $1\la r \la 9\kpc$. The dynamo action at $r\ga 9\kpc$ may be too weak to make the magnetic 
arms observable there, but there are no indications of the magnetic and optical arms crossing at 
or near the co-rotation radius. Nevertheless, the right magnitude of the azimuthal phase shifts
obtained in our models (15--$40^\circ$) is arguably encouraging.

Another possibility is that the spiral structure in NGC~6946
is more complex than that of a single, rigidly rotating pattern.
For instance, the arms may be winding up to some extent, and thus transient \citep[e.g.][]{dobbsetal10, sellwood11, quillenetal11, wadaetal11, grandetal12}.
The model of \citet{comparettaquillen12} invokes multiple rigidly rotating patterns, 
which interfere to produce spiral features that wind up.
Magnetic arms which are present over several kiloparsecs in radius, and which have a large 
negative phase shift from the $\alpha$-arms that varies only weakly with radius, like in NGC~6946,
are indeed found in our `winding-up' spiral model of \citetalias{css12}.

On the other hand, the mutual arrangement of magnetic and material arms is more complicated 
in M51 \citep{PFSBBFH06,fetal11} where the two patterns overlap in some regions but are 
systematically offset by about 0.5--$0.6\kpc$ elsewhere
\citep[adopting $7.6\Mpc$ for the distance to M51 -- ][]{CFJKLD02}.
This linear displacement corresponds to the angular phase difference of 
5--$10^\circ$ at $r=3$--$6\kpc$ where such an offset is best pronounced.
The two arms visible in polarized 
emission at $\lambda6.2\cm$ intersect the material arms observed in other tracers at 
$r=5$--$6\kpc$.
\citet{ESE89}
suggest that M51 has two spiral arm systems, with the co-rotation radii at $r=4.8\kpc$
and $r=12\kpc$.
\citet{GCG93} 
determine the co-rotation radius to be $5.8\kpc$ from CO 
observations and modelling of M51.
Then the region at $r=3$--$6\kpc$ is (mostly) inside the co-rotation radius and the magnetic arms 
in the inner galaxy are displaced downstream of the material arms, as expected in our model.
The two magnetic arms are positioned differently with respect to the material arms at larger 
galactocentric distances. Outside the co-rotation circle, magnetic Arm~1 (on the east of the 
galactic centre) is systematically lagging the material arm, in accordance with our model, but
the magnetic Arm~2 overlaps the material arm.

Our model appears to be better applicable to this galaxy, 
but the diverse mutual arrangement of the magnetic and material arms in M51 strongly suggests that 
more than one physical effect is involved in its genesis.
\citet{dobbsetal10} suggest that the spiral arm morphology in M51 evolves
rapidly due to the interaction with its satellite galaxy. 
Magnetic arms driven by evolving material arms are explored in 
\citetalias[Section~5.1 of][]{css12}. 
It also appears that M51 may be a good candidate for the `ghost' magnetic arms, 
where the magnetic arms trace the spiral arms as they were in the past up to 
a few hundred Myr earlier. 
The simulations of \citet{dobbsetal10} show that the material spiral arms are being wound up by the 
galactic differential rotation, 
so the model with transient material spirals of \citetalias{css12} may be more relevant to M51 
than the model with a rigidly rotating, stationary spiral pattern explored here.

Still another morphology of magnetic and gaseous arms is observed in the nearby barred galaxy M83. 
\citet{BEFPSS03} applied the same approach involving wavelet transforms as \citet{fricketal00} to
determine the positions of the spiral arms as seen in optical light, dust, H$\alpha$, CO, as well 
as the total and polarized intensities at $\lambda\lambda6$ and $13\cm$.
The morphology of M83 is dominated by a well-pronounced two-armed spiral pattern in each tracer, 
with some substructure within the arms. 
The magnetic and material arms in this galaxy are clearly separated but intersect at the galactocentric radius 
$6.4\kpc$ \citep[given the distance to M83 is $4.5\Mpc$ -- ][]{Tetal03}.
The co-rotation radius of M83, determined by \citet{Hetal09} as $2.4$\,arcmin, or $6.2\kpc$, is 
practically equal to the intersection radius given the uncertainties involved. 
This galaxy is a good candidate for the formation of magnetic arms by the mechanism suggested here.

An important assumption of all or most of the available models of magnetic arms is that the 
$\alpha$-effect is stronger within, or at least correlated with, the material spiral arms. 
We note that the effects of the spiral arms on the scale height of the interstellar gas, its turbulent scale 
and velocity, 
local velocity shear, and other parameters that control the intensity of the dynamo action are far 
from being certain, 
either observationally or theoretically \citep[see][and references therein]{sh98,SSNGB04}. 
\citet{SS98} argue that the magnitude
of the $\alpha$ effect within the material arms can be four times smaller than between the arms, 
whereas the turbulent diffusivity is plausible to be only weakly modulated by the spiral pattern. 
As a result, the local dynamo number within the material arms can be four times smaller than 
between them, and the dynamo action can, in fact, 
be more vigorous between the material arms. 
Direct numerical simulations by \citet{elstnergressel12} find that both the $\alpha$ effect and turbulent diffusivity 
increase with star formation rate $\sigma$, which itself will be larger within the arms,
but this leads to an overall dynamo number that scales inversely with $\sigma$.
However, even if the dynamo number is larger in the interarm regions, 
this does not necessarily imply that the steady-state mean magnetic field there should be stronger than within the arms, 
since the gas density is lower in between the arms than within them. 
Significant effort in observations, 
theory of interstellar magnetohydrodynamics and dynamo theory are still required to 
establish a clear understanding of the various mechanisms that produce the
diverse morphology of magnetic arms in spiral galaxies.
Our work here and in \citetalias{css12} provides the beginnings of such a study.
 
\section*{Acknowledgements}
We thank Nishant Singh for useful discussions. We are also grateful to the referee for suggestions that led to significant improvements.
\bibliographystyle{mn2e}
\bibliography{refs_Anv}

\appendix
\section{Details of the asymptotic solutions}
\label{sec:details}
In the no-$z$ approximation, the vertical diffusion terms are approximated as
\[
\frac{\del^2 a_{\pm2}}{\del z^2}\simeq-\frac{\pi^2a_{\pm2}}{4h^2},
\quad 
\frac{\del^2 b_{\pm2}}{\del z^2}\simeq-\frac{\pi^2b_{\pm2}}{4h^2},
\]
where $h$ is the disc half-thickness. Further,
the terms containing $\alpha$ can be approximated as
\begin{align*}
\nonumber\frac{\del}{\del z}(\alp_0 b_0+\alp_2 b_{-2} +\alp_{-2}b_2)&\simeq\frac{2}{\pi h}(\alp_0 
b_0+\alp_2 b_{-2} +\alp_{-2}b_2),\\
\frac{\del}{\del z}(\alp_0 b_{\pm2}+\alp_{\pm2}b_0)&\simeq\frac{2}{\pi h}(\alp_0 
b_{\pm2}+\alp_{\pm2}b_0).
\end{align*}

For the radial diffusion terms, we start by representing 
$a_{\pm2}$ and $b_{\pm2}$ as 
\be
\label{ab}
a_{\pm2}=-\overline{a}(r)\Exp{\mp i\theta_a(r)},\quad b_{\pm2}=\overline{b}(r)\Exp{\mp i\theta_b(r)},
\ee
where the minus sign in front of $\mean{a}$ is introduced for future convenience,
and where we have removed the $z$-dependence of $\overline{a}$ and $\overline{b}$ due to the fact 
that we are now working in the no-$z$ approximation.
Note that $|a_2|=|a_{-2}|$ and $|b_2|=|b_{-2}|$, whereas
the phases of $a_2$ and $a_{-2}$ (or $b_2$ and $b_{-2}$) must be equal in magnitude and opposite in sign, since
$B_r$ and $B_\phi$ are real.
We then apply the WKBJ-type tight-winding approximation, assuming that the amplitude varies much 
less rapidly with radius than the phase, i.e.,
\begin{equation}
\label{tightwinding}
\left|\frac{d\overline{a}}{dr}\right| \ll \left|\overline{a}\frac{d\theta_a}{dr}\right|,
\quad
\left|\frac{d\overline{b}}{dr}\right| \ll \left|\overline{b}\frac{d\theta_b}{dr}\right|.
\end{equation}
The radial diffusion terms are then approximated by
\begin{equation}
\label{rad_diffa}
\frac{d}{dr}\left[\frac{1}{r}\frac{d}{dr}(ra_{\pm2})\right] 
\simeq -\left(\theta'^2_a\pm i\theta''_a\right)a_{\pm2},
\end{equation}
\begin{equation}
\label{rad_diffb}
\frac{d}{dr}\left[\frac{1}{r}\frac{d}{dr}(rb_{\pm2})\right] 
\simeq -\left(\theta'^2_b\pm i\theta''_b\right)b_{\pm2},
\end{equation}
where prime stands for $d/dr$. 
It should be noted that in the region where the relative strength of 
non-axisymmetric components with respect to axisymmetric components is expected to 
be the largest (near the co-rotation radius), we expect $1/r \ll|d/dr|$ since 
the co-rotation radius is typically comparable to the size of the galaxy.
Thus, we have dropped terms proportional to $1/r$.
Since turbulent diffusion terms containing the $\phi$-derivatives are proportional to $1/r^2$, 
they, too, can be neglected.
We find these approximations to be suitable even for the essentially axisymmetric modes that are located much closer to the galactic centre,
where the dynamo number is maximum, 
because even here the radial scale length of the field turns out to be small compared to $r$ in the kinematic regime.

With the above approximations, Eqs.~\eqref{m2a} and \eqref{m2b} yield, after some algebra, 
\be
\label{b2}
b_{\pm2}=b_0\frac{\alpha_{\pm2}}{\alp_0}|D_0|X_{\pm2},
\quad
a_{\pm2}=-Y_{\pm2}b_{\pm2}.
\ee
Here
\be
\label{X}
X_{\pm2}(r)=\frac{(A\mp iB)}{A^2+B^2}=\frac{\Exp{\mp i\beta}}{\sqrt{A^2+B^2}}=X_{\mp2}^*,
\ee
and
\be
\label{Y}
Y_{\pm2}(r)=\Atilde\mp i\Btilde=\sqrt{\Atilde^2+\Btilde^2}\Exp{\mp i\betatilde}=Y_{\mp2}^*,
\ee
with the dynamo number defined as
\be
D_0=\frac{\alpha_0Gh^3}{\eta\turb^2}=\frac{\alpha_0Gt\diff^2}{h}<0,
\ee
where $t\diff=h^2/\eta\turb$ is the vertical turbulent diffusion time scale and
\be
\label{beta}
\cos{\beta}=\frac{A}{\sqrt{A^2+B^2}}, \quad 
\sin{\beta}=\frac{B}{\sqrt{A^2+B^2}},
\ee
with
\be
\label{A}
\bsplit
A&=D_0
+\frac{\pi}{2c_\tau}\Bigg\{t\diff^2(1+\Gam\tau)(\Gam^2-4\omtilde^2)+8\Omega\Gam \omtilde\tau t\diff^2\\
&+\left[(1+\Gam\tau)^2+(2\Omega\tau)^2\right]^{-1}\\
&\times\Bigl[{c_\tau^2(1+\Gam\tau)(\tfrac14\pi^2+\theta_a'^2h^2)(\tfrac14\pi^2+\theta_b'^2h^2)}\\
&\phantom{\times}
-{2c_\tau^2\Omega\tau\left[\left(\tfrac14\pi^2+\theta_a'^2h^2\right)\theta_b''h^2 +\left(\tfrac14\pi^2+\theta_b'^2h^2\right)\theta_a''h^2\right]}\\
&\phantom{\times}
-{c_\tau^2(1+\Gam\tau)\theta_a''\theta_b''h^4}\Bigr]\\
&+c_\tau\Gam t\diff\left(\tfrac12\pi^2+\theta_a'^2h^2+\theta_b'^2h^2\right)
-2c_\tau\omtilde t\diff h^2(\theta_a''+\theta_b'')\Bigg\},
\end{split}
\ee
\be
\label{B}
\bsplit
B&=\frac{\pi}{2c_\tau}
\Bigg\{-2\Omega\tau t\diff^2(\Gamma^2-4\omtilde^2)+4(1+\Gam\tau)\Gam\omtilde t\diff^2\\
&+\left[(1+\Gam\tau)^2+(2\Omega\tau)^2\right]^{-1}\\
&\times\Bigl[2{c_\tau^2\Omega\tau(\tfrac14\pi^2+\theta_a'^2h^2)(\tfrac14\pi^2+\theta_b'^2h^2)}\\
&\phantom{\times}+{c_\tau^2(1+\Gam\tau)\left[(\tfrac14\pi^2+\theta_a'^2h^2)\theta_b''h^2+(\tfrac14\pi^2+\theta_b'^2h^2)\theta_a''h^2\right]}\\
&\phantom{\times}-2{c_\tau^2\Omega\tau\theta_a''\theta_b''h^4}
\Bigr]\\
&+2c_\tau\omtilde t\diff\left(\tfrac12\pi^2+\theta_a'^2h^2+\theta_b'^2h^2\right)
+c_\tau\Gam t\diff h^2(\theta_a''+\theta_b'')\Bigg\}.
\end{split}
\ee
Further,
\be
\label{betatilde}
\cos{\betatilde}=\frac{\Atilde}{\sqrt{\Atilde^2+\Btilde^2}}, \quad 
\sin{\betatilde}=\frac{\Btilde}{\sqrt{\Atilde^2+\Btilde^2}},
\ee
with
\be
\label{Atilde}
\Atilde=-\frac{\Gam}{G}-c_\tau\frac{(1+\Gam\tau)(\tfrac14\pi^2+\theta_b'^2h^2)-2\Omega\tau\theta_b''h^2}{[(1+\Gam\tau)^2+(2\Omega\tau)^2]Gt\diff},
\ee
\be
\label{Btilde}
\Btilde=\frac{2\omtilde}{G}+c_\tau\frac{2\Omega\tau(\tfrac14\pi^2+\theta_b'^2h^2)+(1+\Gam\tau)\theta_b''h^2}{[(1+\Gam\tau)^2+(2\Omega\tau)^2]Gt\diff}.
\ee
Substituting \eqref{alp_n}, \eqref{X} and \eqref{Y} into Eq.~\eqref{b2},
we arrive at
\be
\label{apm2a}
a_{\pm2}=-\frac{|D_0|b_0\epsilon_\alp}{2}\sqrt{\frac{\Atilde^2+\Btilde^2}{A^2+B^2}}
\exp{\left[\mp i(\kappa r+\beta+\betatilde)\right]},
\ee
\be
\label{bpm2a}
b_{\pm2}=\frac{|D_0|b_0\epsilon_\alp}{2}\frac{1}{\sqrt{A^2+B^2}}
\exp{\left[\mp i(\kappa r+\beta)\right]}.
\ee

Using Eqs.~\eqref{ab}, we now make the identifications
\[
\theta_a=\kappa r+\beta+\betatilde, \qquad \theta_b=\kappa r+\beta,
\]
and so
\be
\label{theta_derivs}
\theta'_a=\kappa+\beta'+\betatilde', \quad \theta'_b=\kappa+\beta', \quad \theta''_a=\beta''+\betatilde'', \quad \theta''_b=\beta''.
\ee

Clearly, $\beta$ and $\betatilde$ depend on $A$, $B$, $\Atilde$ and $\Btilde$ (Eqs.~\ref{beta} and \ref{betatilde}), which, in turn, 
depend on the derivatives of $\beta$ and $\betatilde$ through $\theta_a'$, $\theta_b'$, 
$\theta_a''$ and $\theta_b''$ (Eqs.~\ref{A}, \ref{B}, \ref{Atilde} and \ref{Btilde}).
This means that, unless $\beta'$ and $\betatilde'$ are completely neglected, it is not possible to 
obtain an analytical solution.
We will see however that, if the terms involving $\beta'$, $\betatilde'$, $\beta''$ and 
$\betatilde''$ are small compared with the turbulent diffusion terms that do not vary with $r$, 
then $\beta$ and $\betatilde$ can be determined semi-analytically, by successive iteration.

The need for the iterations can be understood as follows.
Firstly, if $\beta'$ and $\betatilde'$ were zero, 
then the radial phase would be $\kappa r$, which is the same as the radial phase of the 
$\alpha$-spiral. Thus, the magnetic arms would be as tightly wound as the $\alpha$-arms.
In fact, the extra contributions to the radial phase $\beta+\betatilde$ for $\mbr$ and $\beta$ for $\mbp$, 
are finite, and turn out to have the same sign as $\kappa$.
This implies that the magnetic arms (defined as the set of positions where $|\meanv{B}|$ is maximum),
are more tightly wound than the material arms that produce them.
This is because the enhancement of field due to the
enhanced dynamo action within the $\alpha$-arm is advected downstream with the 
differentially rotating gas.
Ultimately a balance is set up between the competing effects of preferential 
magnetic field generation in the 
$\alpha$-arms and advection downstream, 
resulting in stationary magnetic arms that are more tightly wound than the $\alpha$-arms, and 
cross them near the co-rotation circle.
Moreover, this balance, which determines the pitch angle of the magnetic arms (ridges), and thus 
the values of $\beta$ and $\betatilde$, 
involves many quantities, including the derivatives of $\beta$ and $\betatilde$ through the radial 
diffusion (as seen in the equations of Appendix~\ref{sec:details}).
This self-dependence results from the fact that the radial diffusion is stronger when the magnetic 
arms (ridges) are more tightly wound, as can be seen from Eqs.~\eqref{rad_diffa} and 
~\eqref{rad_diffb}. 
Thus, the radial diffusion feeds back onto itself, forcing us to iterate to obtain the functions $\beta$ and $\betatilde$.

\section{Growth rates of non-axisymmetric magnetic modes}\label{sec:Gamma}
To obtain the growth rate $\Gamma$,
we substitute Eq.~\eqref{local-global} into the no-$z$ versions of 
Eqs.~\eqref{m0a} and \eqref{m0b}, and, in Eq.~\eqref{m0a}, 
we also substitute Eq.~\eqref{b2} for $b_{\pm2}$. 
In addition, we make use of the relation
\[
\alp_2\alp_{-2}=\alp_0^2\epsilon_\alp^2/4,
\]
obtained from Eq.~\eqref{alp_n}.
Equations~\eqref{m0a} and \eqref{m0b} then reduce to
\[
\begin{split}
(1+\Gam\tau)&\Gam-c_\tau\frac{\eta\turb}{q}\left[\frac{(rq)'}{r}\right]'\\
=&-\frac{2c_\tau}{\pi h}\alp_0 \,\frac{\btilde}{\atilde}\left[1+\frac{|D_0|\eps_\alp^2}{4}(X_2+X_{-2})\right]-c_\tau\eta\turb\frac{\pi^2}{4h^2},
\end{split}
\]
\[
(1+\Gam\tau)\Gam-c_\tau\frac{\eta\turb}{q}\left[\frac{(rq)'}{r}\right]'
	=(1+\Gam\tau)G\frac{\atilde}{\btilde}-c_\tau\eta\turb\frac{\pi^2}{4h^2},
\]
where $X_{\pm2}$ are defined in Eq.~\eqref{X}.
Note that $\eps_\alp=0$ corresponds to an axisymmetric forcing of the dynamo.
It is convenient to introduce a new quantity $\gamma(r)$ defined as
\[
\gamma(r)=(1+\Gam\tau)\Gam-c_\tau\frac{\eta\turb}{q}\left[\frac{(rq)'}{r}\right]',
\]
which, in the $\tau\rightarrow0$ case, 
is equal to the growth rate $\Gamma$ under the local or slab approximation (vanishing radial derivatives).
For this reason, $\gamma$ is sometimes called the `local growth rate'.
In the case of finite $\tau$, $\gamma$ is also closely related to $\Gamma$ in the local approximation
(since typically $\Gamma\tau\ll1$, $\gamma$ can still loosely be thought of as the local growth rate).
Now the equations can be separated into a global eigen-equation for $q(r)$,
\[
(1+\Gam\tau)\Gam q=\gamma q+c_\tau\eta\turb\left[\frac{(rq)'}{r}\right]',
\]
which only contains derivatives with respect to $r$,
and two local equations for $\atilde$ and $\btilde$,
\begin{align*}
\left(\gamma+\frac{\pi^2c_\tau}{4t\diff}\right)\atilde
=&-c_\tau\frac{2\alp_0}{\pi h}\btilde\left[1+\frac{|D_0|\eps_\alp^2}{4}(X_2+X_{-2})\right],\\
\left(\gamma+\frac{\pi^2c_\tau}{4t\diff}\right)\btilde
=&(1+\Gam\tau)G\atilde,
\end{align*}
which generally only have derivatives with respect to $z$ but become
algebraic equations under the no-$z$ approximation.

\label{lastpage}
\end{document}